%% file: main.tex
\documentclass[submission,copyright,creativecommons]{eptcs}
\providecommand{\event}{SYNT 2017} 
\usepackage{breakurl}              
\usepackage{underscore}            
\usepackage{mathrsfs}
\usepackage{amsmath}
\usepackage{amssymb}

\usepackage{wrapfig}
\usepackage{subfigure}
\usepackage{tikz}
\usepackage{color}
\usepackage{xcolor}
\usepackage[draft]{todonotes}    
\usepackage{ulem}
\usepackage{bytefield}

\usepackage{longtable}

\usepackage{array}
\newcolumntype{H}{>{\setbox0=\hbox\bgroup}c<{\egroup}@{}}

\usetikzlibrary{arrows,petri,backgrounds, positioning, shapes, patterns,calc}
\tikzstyle{envplace}=[circle,thick,draw=black!75,fill=white,minimum size=6mm]
\tikzstyle{sysplace}=[circle,thick,draw=black!75,fill=black!20,minimum size=6mm]
\tikzstyle{transition}=[rectangle,thick,draw=DarkBlue!75,fill=DarkBlue!20,minimum size=4mm]
\tikzstyle{bad}=[double=black!20]
\tikzstyle{badstate}=[pattern=checkerboard, fill opacity=0.1, draw opacity=1, draw=black, text opacity=1]
\tikzstyle{win}=[]

\definecolor{DarkBlue}{rgb}{0.1,0.1,0.5}
\newcommand{\DarkBlue}{\color{DarkBlue}}
\definecolor{DarkRed}{rgb}{.7,0.5,0.5}

\definecolor{Red}{rgb}{.7,0,0}
\newcommand{\Red}{\color{Red}}
\definecolor{ganttGreen}{HTML}{2a8728}
\definecolor{cdcBlueL}{rgb}{0.5,0.589,0.793} 
\definecolor{cdcGreenL}{rgb}{0.688,0.828,0.605} 

\newcommand{\adam}{\textsc{Adam}}

\newcommand{\defEQ}{\stackrel{\mathit{Def.}}{=}}

\newcommand{\pl}{\mathcal{P}}
\newcommand{\plS}{\pl_S}
\newcommand{\plE}{\pl_E}
\newcommand{\tr}{\mathcal{T}}

\newcommand{\preset}[1]{{}^\bullet #1}
\newcommand{\postset}[1]{{#1}^\bullet}
\newcommand{\type}{\mathtt{type}}

\newcommand*\circled[1]{\tikz[baseline=(char.base)]{
            \node[shape=circle,draw,inner sep=1pt] (char) {\bf\tiny#1};}}

\title{Symbolic vs.\ Bounded Synthesis for Petri Games\thanks{This work was partially supported by the European Research Council (ERC) Grant OSARES (No.\ 683300) and by the German Research Foundation through the Research Training Group (DFG GRK 1765) SCARE.}}
\author{
Bernd Finkbeiner
\institute{Universit\"at des Saarlandes\\ Saarbr\"ucken, Germany}
\email{finkbeiner@react.uni-saarland.de}
\and
Manuel Gieseking
\institute{Carl von Ossietzky Universit\"at Oldenburg\\ Oldenburg, Germany}
\email{gieseking@informatik.uni-oldenburg.de}
\and
Jesko Hecking-Harbusch
\institute{Universit\"at des Saarlandes\\ Saarbr\"ucken, Germany}
\email{hecking-harbusch@react.uni-saarland.de}
\and 
Ernst-R\"udiger Olderog
\institute{Carl von Ossietzky Universit\"at Oldenburg\\ Oldenburg, Germany}
\email{olderog@informatik.uni-oldenburg.de}
}
\def\titlerunning{Symbolic vs.\ Bounded Synthesis for Petri Games}
\def\authorrunning{B.\ Finkbeiner, M.\ Gieseking,{\ ~}\\J.\ Hecking-Harbusch, \& E.-R.\ Olderog} 

\begin{document}
\maketitle

\begin{abstract}
Petri games are a multiplayer game model for the automatic synthesis of distributed systems. We compare two fundamentally different approaches for solving Petri games. The symbolic approach decides the existence of a winning strategy via a reduction to a two-player game over a finite graph, which in turn is solved by a fixed point iteration based on binary decision diagrams~(BDDs). The bounded synthesis approach encodes the existence of a winning strategy, up to a given bound on the size of the strategy, as a quantified Boolean formula (QBF). In this paper, we report on initial experience with a prototype implementation of the bounded synthesis approach. We compare bounded synthesis to the existing implementation of the symbolic approach in the synthesis tool \adam{}. We present experimental results on a collection of benchmarks, including one new benchmark family, modeling manufacturing and workflow scenarios with multiple concurrent processes.
\end{abstract}


\section{Introduction}

The \emph{synthesis problem} asks, given a formal specification, for the existence of an implementation and derives it automatically if existent~\cite{church1963application}. This problem can be described as a game between the system and the environment. A strategy is winning for the system and therefore corresponds to an implementation if the strategy fulfills the winning condition of the game against all behaviors of the environment. The synthesized implementation for a given specification is correct by construction, which is beneficial for error-prone implementation tasks. Synthesis has been applied successfully to implement several practical applications like the protocol of the AMBA bus circuit~\cite{DBLP:conf/date/BloemGJPPW07}. 

Synthesis is especially interesting for \emph{distributed systems} where several concurrent components can communicate with each other to fulfill the specification. Pnueli and Rosner defined the first setting for distributed synthesis~\cite{DBLP:conf/popl/PnueliR89}. \emph{Information forks} have been identified as a necessary and sufficient criterion for the undecidability of the synthesis problem in this setting~\cite{DBLP:conf/lics/FinkbeinerS05}, which prevents most practical applications. Even for the decidable cases without information forks like pipelines and rings, the synthesis problem has non-elementary complexity~\cite{DBLP:conf/focs/PnueliR90,DBLP:conf/lics/KupfermanV01}. More recently, Zielonka's asynchronous automata were introduced as another framework for distributed synthesis. Whereas the synthesis problem for some cases has non-elementary complexity, the general complexity of synthesis for asynchronous automata remains open~\cite{DBLP:journals/ita/Zielonka87,DBLP:conf/fsttcs/MadhusudanTY05}.

We take Petri games as a starting point to synthesize distributed systems~\cite{DBLP:journals/iandc/FinkbeinerO17}. Petri games define a multiplayer game model where several distributed system players cooperatively play against several distributed environment players. Petri games are based on an underlying Petri net where each token represents a player. Each player in a Petri game has only local information about other players it synchronized with on joint transitions.

We compare two fundamentally different approaches to synthesize winning strategies of Petri games: The \emph{symbolic} approach is based on the result that the synthesis problem for Petri games with a bounded number of system players, a single environment player, and the avoidance of bad places as winning condition can be solved in single exponential time~\cite{DBLP:journals/iandc/FinkbeinerO17}. This approach performs a reduction to a two-player game over a finite graph with full information. The implementation is realized in the tool \adam{} using a BDD-based fixed point iteration~\cite{DBLP:conf/cav/FinkbeinerGO15}. The restriction to only a single environment player is an impediment for convenient modeling.

The second approach to find winning strategies in a Petri game is \emph{bounded} synthesis~\cite{DBLP:journals/sttt/FinkbeinerS13}. Here, the size of possible strategies and of the proof that the strategy is winning is limited. It is increased incrementally when no strategy of the previous size can be found. This steers the search towards small winning strategies at the cost of being unable to prove the non-existence of a winning strategy. Bounded synthesis can find winning strategies for Petri games with more than one environment player~\cite{DBLP:conf/birthday/Finkbeiner15}. The bounded approach for Petri games limits the number of fired transitions in the proof of the strategy being winning. It further takes a second bound on the size of the memory each player can utilize as the memory of players is infinite in general. For the pair of bounds, bounded synthesis searches for termination or loops in the strategy and can thereby prove the existence of an infinite winning strategy. This search is encoded in a quantified Boolean formula~(QBF) and solved by a QBF solver. We report on our experience with a first prototype implementation of bounded synthesis generating the QBF and solving it with the QBF solver \textsc{QuAbS}~\cite{conf/sat/Tentrup16} in comparison to the symbolic approach implemented in \adam{}.

\adam{} has been used to synthesize several case studies from manufacturing and workflow scenarios. These case studies have been extended into scalable benchmark families to evaluate the behavior of the implementation of the symbolic approach and to show the applicability of Petri games to synthesize distributed systems~\cite{DBLP:conf/cav/FinkbeinerGO15}. 


The key contributions of this paper are the following:
\begin{itemize}
	\item We add the new benchmark family of a distributed \emph{alarm system} to the benchmark families of \adam{}. The new benchmark family serves to secure several locations with an alarm system such that a burglary at any location is indicated at all locations including the information where exactly the burglary takes place.
	\item We state our experience with the prototype implementation for the generation of the QBFs representing the bounded synthesis problem for the given pair of bounds and with the solving of the generated QBFs. We found out that the bounded unfolding of Petri games benefits from pruning techniques and that solvers for non-CNF QBFs refining an abstraction based on counterexamples show the best performance for solving.
	\item We empirically compare the symbolic synthesis approach implemented in \adam{} with the bounded synthesis approach realized by our prototype implementation. The symbolic approach solves more instances overall from the extended set of benchmark families whereas bounded synthesis derives strategies of smaller size. We present reasons for the observed behavior based on the number of variables in the two approaches.
\end{itemize}
The remainder of the paper is structured as follows. Section~\ref{sec:recap} recaps the theory of Petri games on an abstract level and introduces the distributed alarm system as a running example. In Sec.~\ref{sec:solving}, the symbolic and the bounded approach for solving Petri games are presented, including their application to the distributed alarm system. The experimental results comparing the two approaches are given in Sec.~\ref{sec:expResults}. 

\section{Petri Games}
\label{sec:recap}

Petri games are a multiplayer game model for the synthesis of distributed systems~\cite{DBLP:journals/iandc/FinkbeinerO17}. They define a game on an underlying Petri net by characterizing certain places as bad such that the system has to cooperate to avoid reaching these places to win. The distinction between system and environment is achieved by distributing each place of the Petri net to either belong to the system or to the environment. System places are depicted gray whereas environment places remain white. The players in a Petri game are the tokens flowing through the underlying Petri net. If a token resides in a system place then it is controlled by the token's strategy whereas the behavior of tokens in environment places is uncontrollable. Each token in a Petri game has local memory, which is only exchanged with the other participating tokens of joint transitions. The causal history of tokens is utilized by their local strategy to make decisions on which transitions to fire. Therefore, the strategy of the system in a Petri game is a local controller for each process and not a global controller with information about the state of tokens in all system places.
	
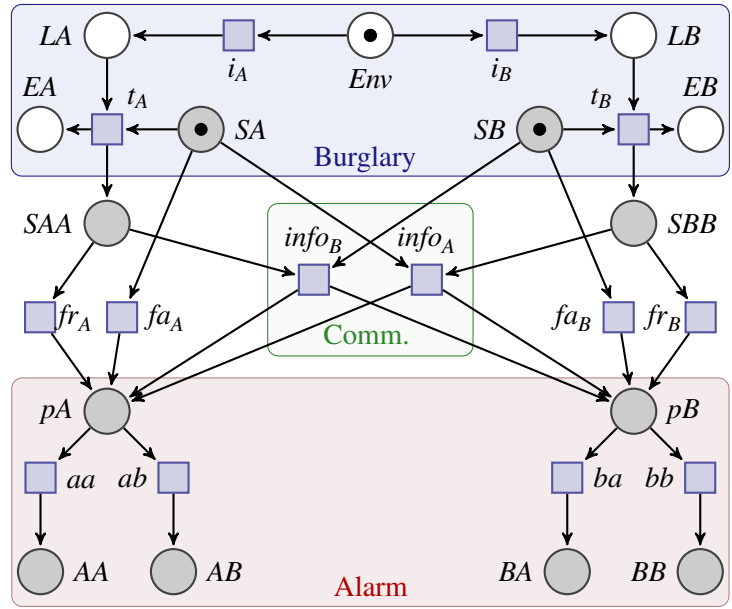
\begin{figure}
 \centering
 \input{./figures/securitySystem}
 \caption{\it{} The Petri game depicting the synthesis problem of an alarm system being distributed over two independent locations with the possibility to communicate. The alarm system has to detect a burglary and set off alarms at both locations.}
  \label{fig:SecSystem}
\end{figure}
\noindent
{\em
{\bf{}Example.} 
We model a distributed alarm system.
A burglar (modeled by the environment player residing in place \(\mathit{Env}\)) 
decides to intrude one of several secured locations. 
In Fig.~\ref{fig:SecSystem}, the situation is displayed
for the distributed locations \(A\) (shown to the left) 
and \(B\) (shown to the right). 
This situation has been used as the introductory example in~\cite{DBLP:journals/iandc/FinkbeinerO17}
and is extended to a benchmark family for an arbitrary number of secured locations later in this paper.
The goal of the game is to find a strategy for each of the two alarm processes
initially residing in places \(\mathit{SA}\) and \(\mathit{SB}\).
A token in place \(\mathit{XY}\) \emph{(}for \(X,Y \in\{A,B\}\)\emph{)} represents 
that in location~\(X\) an alarm is set off indicating that 
the strategy of the alarm system presumes an intrusion at location \(\mathit{Y}\).
If the burglar intrudes location \(A\) by entering place~\(\mathit{LA}\),
the strategies should steer the token in \(\mathit{SA}\) to place 
\(\mathit{AA}\) and the token in \(SB\) to place \(\mathit{BA}\) 
in order to correctly indicate the burglar's intrusion,
and similarly for the burglar intruding location~\(B\).
If a token resides in one of the system places \(\mathit{SA}\), \(\mathit{SB}\),
\(\mathit{SAA}\), \(\mathit{SBB}\), \(\mathit{pA}\), or \(\mathit{pB}\) 
then the strategy of the player has to resolve nondeterminism between the outgoing transitions.
For this, the strategies for the players in \(\mathit{SA}\) and \(\mathit{SB}\) 
need to collect sufficient information
about the moves of the other system player and the burglar.
For example, the player in place \(\mathit{SB}\) does not know whether the
alarm system in~\(\mathit{SA}\) has fired transition \(t_A\) unless it gets
informed by the communication transition \(\mathit{info}_B\).
We omit the bad places and transitions to them representing false alarms and false reports, 
which have to be avoided by a correct alarm system.
A false alarm occurs when an intrusion is indicated before the burglar actually intruded the object
whereas a false report occurs when the alarm system at a certain location indicates an intrusion
at a location where no intrusion occurred.
}

Formally, a Petri game $\mathcal{P} = (\mathcal{P}_S , \mathcal{P}_E , \mathcal{T} , \mathcal{F} , \mathit{In} , \mathcal{B} )$ is based on a Petri net $\mathcal{N} = (\mathcal{P} , \mathcal{T} , \mathcal{F} , \mathit{In} )$ with the set of places $\mathcal{P} = \mathcal{P}_S \cup \mathcal{P}_E$, the set of transitions $\mathcal{T}$, the flow relation $\mathcal{F} \subseteq (\mathcal{P} \times \mathcal{T}) \cup (\mathcal{T} \times \mathcal{P})$, and the initial marking~$\mathit{In}$. A Petri game divides the places to either belong to the system~($\mathcal{P}_S$) or to the environment~($\mathcal{P}_E$) and it further defines some places as bad ($\mathcal{B} \subseteq \mathcal{P}$). We restrict ourselves to \emph{safe} Petri games, i.e., at most one token can reside in each place. We identify a token residing in a place by the name of the place. A \emph{marking} is a set of places where one token resides at each place. From a marking, a transition is \emph{enabled} if all places in the transition's preset have one token residing in them. The \emph{firing} of an enabled transition removes one token from each place in the transition's preset~($\preset{t}$) and creates one token in each place in the transition's postset~($\postset{t}$). We also refer to the sets of transitions preceding and following a place as the place's preset ($\preset{p}$) and postset ($\postset{p}$). The behavior of Petri games is defined by sequences of \emph{reachable markings}. These sequences start from the initial marking and between each pair of subsequent markings one transition is fired. We say that a place \(p\) is \emph{reachable} if there exists a sequence of reachable markings such that one of the markings contains \(p\).

The notion of \emph{unfoldings} from Petri nets~\cite{DBLP:series/eatcs/EsparzaH08} translates to Petri games. In an unfolding, all cycles, i.e., all reachable sequences of markings starting and ending with the same marking, are unrolled infinitely and all backward branches, i.e., places with at least two transitions merging into them, are unfolded. A place is unfolded by copying the place and all its outgoing transitions including the following sub-games and changing one of the incoming flows of a transition from the place's preset to the copied place. Therefore, the unfolding explicitly represents the unique causal history of each process. As for Petri nets, the unfolding can be infinite for Petri games. In \emph{bounded unfoldings} of Petri games~\cite{DBLP:conf/birthday/Finkbeiner15}, loops and backward branches are only unfolded up to a given bound~$b$ for the number of copies per place. Upon reaching the bound for a place, both the original place and its existing copies can be part of a loop or can have backward branch. Therefore, the bounded unfolding is finite even for Petri games with infinite unfoldings. One bounded unfolding for the Petri game in Fig.~\ref{fig:SecSystem} is depicted in Fig.~\ref{fig:SecSystemBoundedUnfolding} in Sec.~\ref{sec:bounded}. The places $\mathit{AA}$, $\mathit{AB}$, $\mathit{BA}$, and $\mathit{BB}$ are not unfolded despite them being reachable with different causal pasts, i.e., having backward branches.

A \emph{strategy} of a Petri game is a restriction to the unfolding of the Petri game where certain branches of transitions and places are removed. We assume that this removal is based on the decisions of system players not to fire certain transitions. The behavior of a strategy is defined by the remaining sequences of reachable markings. The following four requirements have to hold for a strategy to be \emph{winning}:
\begin{enumerate}
	\item \emph{Safety.} No marking containing a bad place is reachable in the strategy.
	\item \emph{Determinism.} For all system places in all reachable markings of the strategy, at most one outgoing transition is enabled.
	\item \emph{Deadlock avoidance.} For all reachable markings in the strategy, if a transition is enabled for that marking in the unfolding then one transition is enabled in the strategy as well. This requirement prevents trivial solutions where the system does not fire any transition to avoid reaching a marking containing a bad place.
	\item \emph{Justified refusal.} When a transition is removed from the unfolding to create the strategy then there is at least one system place in the removed transition's preset for which all copies of the removed transition are removed as well. This prevents the system from differentiating copies of transitions resulting from the unfolding.
\end{enumerate}

Petri games with a bounded number of system tokens, one environment token, and bad places as winning condition can be solved in single exponential time~\cite{DBLP:journals/iandc/FinkbeinerO17}. 

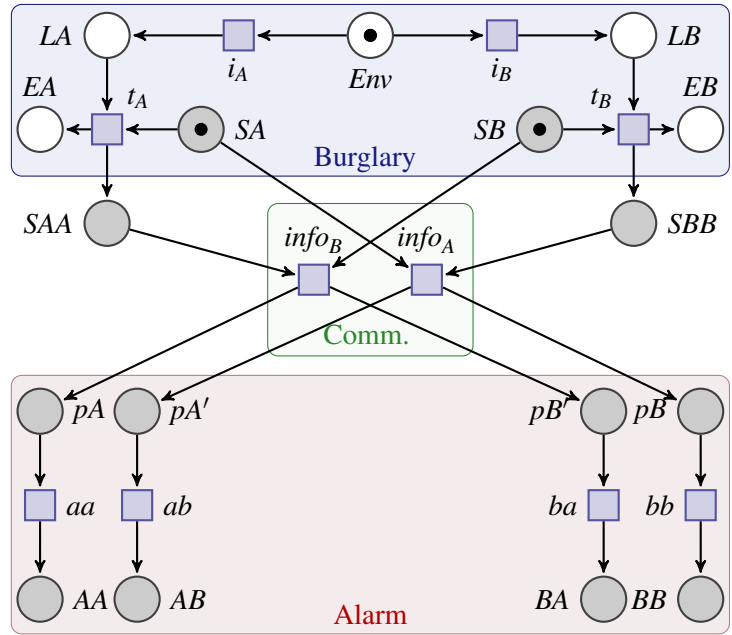
\begin{figure}
 \centering
 \input{./figures/securitySystemStrat}
 \caption{\it{} The winning strategy for the system players for the Petri game depicted in Fig.~\ref{fig:SecSystem} modeling an alarm system. To win, both system players have to communicate the detection of the burglary before setting off their alarms.}
  \label{fig:SecSystemStrat}
\end{figure}

\begin{figure}
 \centering
 \input{./figures/securitySystemDistrStrat}
 \caption{\it{} The distributed local controllers for each player of the winning strategy depicted in Fig.~\ref{fig:SecSystemStrat}. The parallel bars indicate the parallel composition of Petri nets \cite{olderog91} for the environment and the two system controllers, with synchronization on equally labeled transitions. First, the burglary is detected, then the detection is communicated, and afterwards the alarm is set off.}
  \label{fig:SecSystemDistrStrat}
\end{figure}
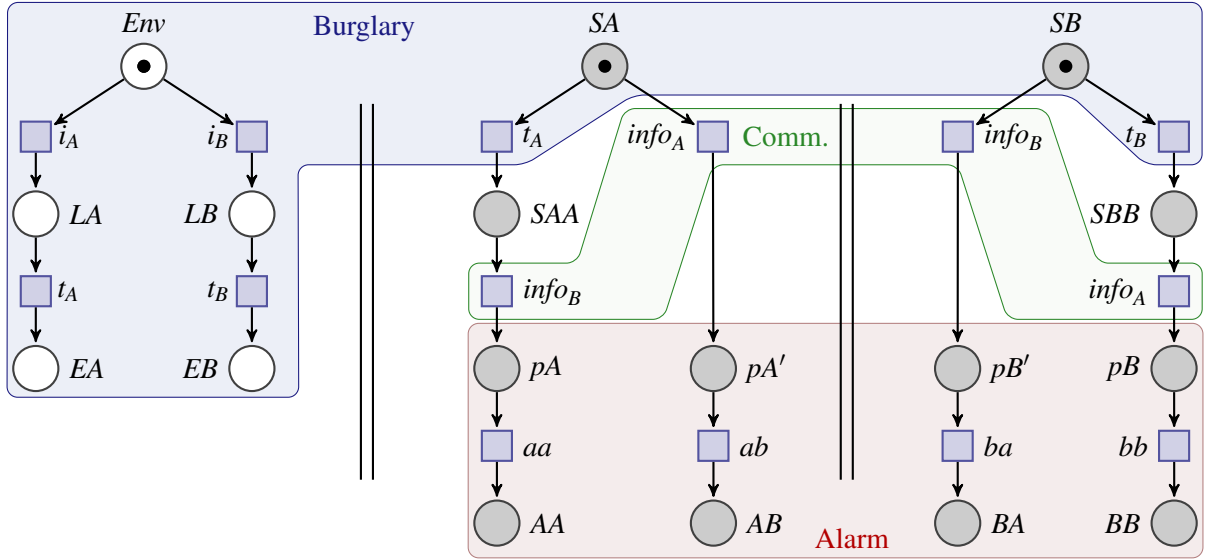

{\em{\bf{}Example.}
The winning strategy for the alarm system from Fig.~\ref{fig:SecSystem} is depicted in Fig.~\ref{fig:SecSystemStrat}. The displayed strategy avoids bad places because no false alarm can occur, as the alarm is always triggered after the burglar intruded, and because no false report can occur, as both system tokens exchange information and utilize it to indicate the correct location of intrusion. The strategy is deterministic because only one of the two respective outgoing transitions of $\mathit{SA}$ and $\mathit{SB}$ is enabled, depending on the location of intrusion, and all other system places have only one outgoing transition. The strategy is deadlock-avoiding because after indicating the alarm, the system terminates as no transitions are enabled. The unfolding of the game only allows justified refusal by the system. Therefore, the displayed strategy is winning.

The strategy contains the local controllers for the alarm systems at location $A$ and at location $B$. The local controller for the alarm system at $A$ behaves in the following way. The alarm system at~$\mathit{SA}$ waits until it either recognizes an intrusion at $A$ via transition $t_A$ or is informed about an intrusion at $B$ via transition $\mathit{info}_A$. The place $\mathit{pA}'$ is reached after getting informed about an intrusion at $B$ from which the transition $\mathit{ab}$ is fired setting off an alarm at $A$ indicating an intrusion at $B$. The place $\mathit{SAA}$ is reached after recognizing an intrusion at $A$. The local alarm system informs the local alarm system at $B$ with the transition $\mathit{info}_B$ and afterwards fires the transition $\mathit{aa}$ to reach the place $\mathit{AA}$. This place represents an alarm at location $A$ that there was an intrusion at location $A$. The two local system controllers can be found in Fig.~\ref{fig:SecSystemDistrStrat}. Note that they can only behave correctly as they rely on the other local alarm system to faithfully inform them about an ongoing burglary, i.e., the system players cooperate. This distribution of a winning strategy into local controllers is possible for all winning strategies of safe, concurrency-preserving Petri games with only one environment player~\cite{DBLP:journals/iandc/FinkbeinerO17}.
}

A \emph{bounded strategy} for a bounded unfolding is generated in the same way as a strategy is generated for the unfolding. Based on the decisions of system players, transitions and the following sub-games are removed. The bounded strategy has to fulfill the same requirements as a strategy for an unfolding but may have fewer system places where transitions can be removed. Places are not unfolded infinitely often in the bounded unfolding, which implies that certain histories are aggregated in one place for which the same decision is repeated. A bounded strategy for a bounded unfolding can easily be extended into a strategy for the general unfolding by repeating the same decisions at places, which were not unfolded in the bounded unfolding. The converse direction is not true, i.e., a strategy for the unfolding cannot in general be translated into a strategy for a bounded unfolding~\cite{DBLP:conf/birthday/Finkbeiner15}.

\section{Symbolic and bounded solving}
\label{sec:solving}
We recall the symbolic solving approach implemented in \adam{} and the bounded solving approach implemented as a prototype. Both approaches are compared in Sec.~\ref{sec:expResults}. Symbolic solving is based on the reduction of Petri games to a two-player game on finite graphs. This is solved using BDDs in \adam{}. The bounded solving approach utilizes two bounds $n$ on the size of the proof and $b$ on the available memory. The question of the existence of a strategy within these bounds is encoded in a 2-QBF. Our prototype automates the generation of the 2-QBF, invokes the QBF solver \textsc{QuAbS} to solve it, and generates a winning strategy if the QBF is satisfiable.

\subsection{Symbolic Solving}
\label{sec:unbounded}

In the symbolic synthesis approach for Petri games introduced in~\cite{DBLP:journals/corr/FinkbeinerO14}, the existence of a winning strategy for the system players is decided via a reduction to a two-player game over a finite graph with complete information. We recap the ideas of the reduction in this section for the comparison to the bounded approach presented in Sec.~\ref{sec:bounded}. 
In this paper, we only consider the case of one environment and a bounded number of system players for the symbolic synthesis approach.
This case can be solved with a safety objective in single-exponential time~\cite{DBLP:journals/iandc/FinkbeinerO17}.
Furthermore, we stick to safe Petri nets, 
since the implementation of \adam{} and the bounded synthesis approach are limited to safe nets.

The general approach for the symbolic solving of Petri games is done in three steps: Firstly, from a given safe Petri game with
one environment player, a bounded number of system players, and a safety objective, a two-player game over a
finite graph is constructed. The environment player is represented by Player 1 (depicted as white rectangles with sharp corners) and all system players together are represented
by Player 0 (depicted as gray rectangles with rounded corners). Secondly, a winning strategy of the two-player game is constructed such that the system players can cooperatively
play against all behaviors of the hostile environment without encountering any bad situations. Thirdly, the winning strategy
for the system players (Player 0) of the two-player game over a finite graph is transformed into a winning strategy of the system players in the Petri game
and distributed into local controllers for each system player.

The two-player game over a finite graph simulates a subset of the behavior of the Petri game in such a way that the game over the graph
can be considered as \emph{completely informed}, i.e., both players have full information about their opponent at all times.
Even though only a subset of the 
behavior is considered, \cite{DBLP:journals/iandc/FinkbeinerO17} shows the existence of a strategy of the Petri game if
and only if a strategy for the two-player game exists. Intuitively, the omitted behavior corresponds to situations where
the system players exploit knowledge about the environment player's behavior of which they had not been informed.
The key idea to achieve complete information is to delay every \emph{environment transition}, i.e., transitions
\(t\in\tr\) with \(\preset{t}\cap{\plE}\neq\emptyset\), until every next possible action of the system has to be 
done directly or indirectly in interaction with the environment (or there is no future interaction with the environment needed at all).
Those states of the two-player game where the system players have progressed maximally are called \emph{mcut}s.
In an mcut, all system players will be informed of the environment's decision when executing their next step (or they will never be informed of the environment's decision). 
This idea restricts the proposed solving technique to only one environment player.
The states corresponding to an mcut are assigned to the environment (Player~1) and all other states to the system (Player~0).

To simulate the Petri game, the states of the two-player game correspond to \emph{cuts}, i.e., maximal sets of concurrent
places. For the sophisticated handling of the causal memory model of Petri games, each place of a cut is enriched by a \emph{commitment set},
i.e., a set of transitions currently selected by the corresponding system player to be allowed to fire. The 
transition relation of the two-player game mimics the firing of \emph{chosen} (and enabled) transitions of the Petri net between the corresponding cuts.

Additionally, there is one extra kind of transitions in the two-player game allowing the system players to chose new commitment sets. 
Therefore, each place in a state is equipped with a Boolean flag \(\top\). It is set to true for a place \(p\) in a state \(s\)
if and only if \(s\) is a successor of an mcut reached by firing transition \(t\) and \(p\in\postset{t}\) holds.
Note that it is only harmful for successors of mcuts to directly choose their commitment sets without such an 
intermediate state with a true \(\top\)-flag, since for a winning strategy of the system players, \emph{all} successors of the
environment states have to avoid bad situations. Thus, the commitment sets of system state successors are directly chosen.
The resolving of the \(\top\), i.e., choosing new commitment sets, has to be made before any transition is allowed to fire
to ensure the correct modeling of the players' informedness.
It is therefore guaranteed that every decision of the system, which should be independent of the environment's decision, is actually taken independently. 


Since environment decisions are delayed until the system players have maximally progressed, possibly infinite 
calculations can be encountered when the system players can infinitely proceed without any interaction with the environment.
To prevent these infinite behaviors, a further Boolean flag \(\type_2\) for each place in a state is introduced.
This flag set to true prohibits the corresponding player to maximally progress. Thus, in an mcut all non-\(\type_2\)-typed
places are blocked until the environment makes its next move due to the non-existence of an enabled and chosen transition and the \(\type_2\)-typed
places are blocked by definition. This ensures that the system players cannot pass over the environment's decision by just playing infinitely on their own.

There are three different types of bad situations in the two-player game. Firstly, a state \(s\) is bad if two different transitions \(t_1\)
and \(t_2\) with \(\preset{t_1}\cap\preset{t_2}\neq\emptyset\) are enabled and chosen in \(s\). Those situations are
called \emph{nondeterministic}. Secondly, a state represents a bad situation if it contains bad places.
Thirdly, \emph{deadlock}s are bad situations. Deadlocks are states \(s\) where a transition exists which is enabled in the
corresponding cut of the underlying Petri net, but there is no enabled and chosen transition in \(s\).
For more details, we refer the reader to~\cite{DBLP:journals/corr/FinkbeinerO14,DBLP:conf/cav/FinkbeinerGO15,DBLP:journals/iandc/FinkbeinerO17}.


\begin{figure}
 \centering
 \input{./figures/securitySystemGG}
 \caption{\it{}The part of the two-player game over a finite graph constructed from the Petri game depicted in Fig.~\ref{fig:SecSystem}. The states belonging to the environment player are white with sharp corners whereas the states belonging to all system players together are gray with rounded corners. The winning strategy for the system players is underlaid in orange. Dashed states indicate that not all possible successors are depicted. Checkerboard patterned states designate bad states of the two-player game.}
  \label{fig:SecSystemGG}
\end{figure}
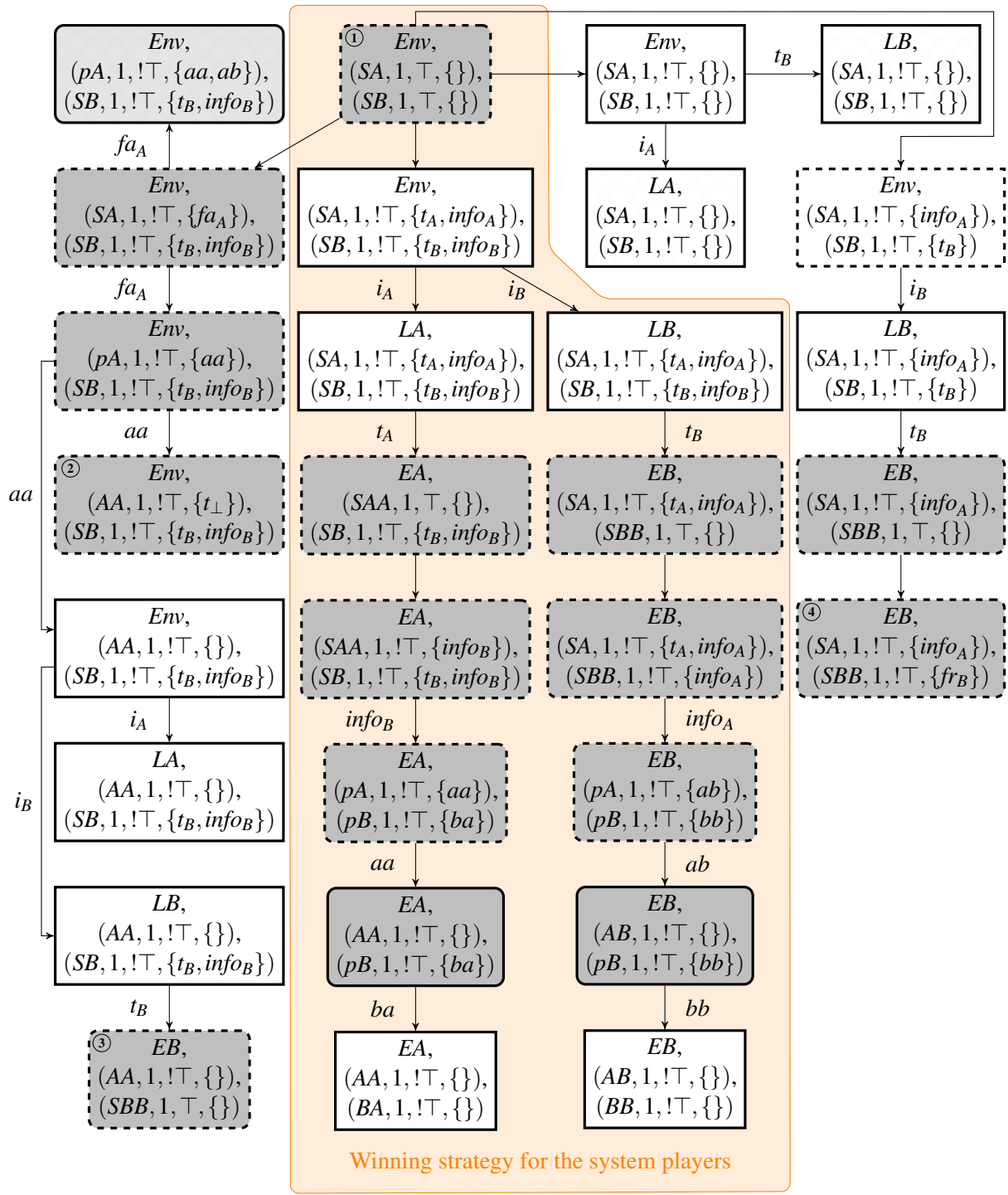
\noindent
{\em
{\bf{}Example.} 
We describe the two-player game over a finite graph obtained by the reduction of the Petri game of Fig.~\ref{fig:SecSystem}.
In Fig.~\ref{fig:SecSystemGG}, a part of this game is visualized. Each state \(s\) is depicted as tuple
\(s\in\plE\times\mathbb{P}\left(\plS\times\{0,\,1\}\times\{\top,\,!\top\}\times\mathbb{P}(\tr)\right)\), where \(0\) indicates
that the corresponding place is \(\type_2\) typed. In this example, there is no possibility for the system players to play
infinitely long without any further interaction with the environment. Thus, all places are typed as not \(\type_2\) (indicated by \(1\)).

The initial state (labeled with 1) corresponds to the cut of the initial marking of the Petri net, where all system tokens initially
have to decide on their commitment sets. A commitment set for a place \(p\) has to be chosen from the powerset of \(\postset{p}\).
Hence, for the initial state the player on \(\mathit{SA}\)
chooses from \(\mathbb{P}(\{\mathit{t_A},\,\mathit{fa}_A,\,\mathit{info}_A\})\) and \(\mathit{SB}\)
from \(\mathbb{P}(\{\mathit{t_B},\,\mathit{fa}_B,\,\mathit{info}_B\})\). All of these possible combinations yield a successor 
of the initial state. Here, only four successors are displayed. 
In general, dashed borders designate that not all successors of a state are depicted in this figure.


The checkerboard patterned states designate the bad situations of the game. Consider for example the upper left state.
There, the token in \(\mathit{pA}\) has two possible chosen and enabled transitions (\(\mathit{aa}\) and \(\mathit{ab}\)).
This situation corresponds to a nondeterministic choice in the Petri game.
The other two depicted bad states correspond to deadlock situations, since the players in \(\mathit{SA}\)
and \(\mathit{SB}\) decided not to allow to fire any transition, but the underlying Petri net can still fire in the
corresponding cut \(\{\mathit{LA},\mathit{SA},\mathit{SB}\}\) (e.g., transition \(t_A\) is enabled).
The situations where the game enters a state containing a bad place are not directly visualized, but reaching a bad
place cannot be prevented in both situations representing a false alarm (depicted as the left branch in Fig.~\ref{fig:SecSystemGG})
and a false report (depicted as the right branch).
In the depicted case of a false alarm, the alarm system of location \(A\) decides to use transition \(\mathit{fa}_A\)
and since the environment is delayed until all system behavior is maximally processed, the alarm system will show a burglary
before it has happened (cf.\ state 2 with a transition \(t_\bot\) leading to a bad place and
\(\preset{t_\bot}=\{\mathit{Env},\mathit{AA}\}\)). Even if the system decides to not use any
of the bad transitions, it cannot avoid a bad situation because it will end up in a deadlock
(cf.\ state 3, where reaching a deadlock is mandatory).
This is similarly in the depicted case of a false report. Since alarm system \(B\) decided to use transition \(\mathit{fr}_B\) 
(cf.\ state 4)
and thus does not report the burglary at its location to the alarm system in \(A\), the token in \(\mathit{SA}\) triggers a deadlock.
If \(\mathit{SA}\) would have chosen some of the other possible transitions
(\(t_A\) or \(\mathit{fa}_A\)), it would still have been a deadlock or a false alarm. The only possible solution for the
alarm systems is to wait for the burglary and then use their information channel (\(\mathit{info}_A\) and \(\mathit{info}_B\))
to inform the other player of the burglary. This situation corresponds to the orange underlaid states resulting in a winning strategy.
}


\subsection{Bounded Solving}
\label{sec:bounded}

We recall the bounded synthesis approach for Petri games~\cite{DBLP:conf/birthday/Finkbeiner15}. In bounded synthesis, a bound is introduced to limit the search space of possible winning strategies to small strategies. Therefore, bounded synthesis can find small implementations fast. The bound is increased incrementally if no winning strategy can be found. If a winning strategy for a certain bound is found, bounded synthesis ensures that this solution is winning in general. We denote this bound by $n$. Bounded synthesis constitutes a semi-decision procedure, i.e., it can prove the existence of a winning strategy but not the non-existence of winning strategies in general. Bounded synthesis finds strategies, which are, because of their small size, interesting for practical applications as they avoid unnecessary (and possibly expensive) computation steps.

In Petri games, each local player can have different information about the other players. Recall that only participating players of a fired transition exchange their complete causal history. A place can have infinitely many different histories, which are represented explicitly in the possibly infinite unfolding. For bounded synthesis of Petri games, we have to introduce a second bound $b$ on the size of the memory for each place in order to retain a finite representation of the bounded synthesis problem. A player residing in the place can only differentiate causal history up to this bound and further history is treated on par with some previous history. The original bound of bounded synthesis $n$ limits the size of the proof of correctness for the strategy. It defines how many transitions are fired until the game has to terminate or has to repeat its behavior in a loop while fulfilling the requirements for a winning strategy. The conditions for a winning strategy are safety, determinism, deadlock-avoidance, and justified refusal as discussed in Sec.~\ref{sec:recap}.

We utilize 2-QBFs to realize the bounded synthesis approach for Petri games. 2-QBFs restrict QBFs to only have one quantifier alternation. A QBF starts with an alternation of either existentially~($\exists$) or universally~($\forall$) quantified sets of Boolean variables. This prefix is followed by the matrix which is a Boolean formula using the standard operators ($\wedge$, $\vee$, $\neg$) and abbreviated operators ($\implies$, $\iff$) on Boolean variables. We focus on 2-QBFs of the form $\exists\, V_1\ldotp \forall\, V_2\ldotp \phi$ where $V_1 \cup V_2$ are all Boolean variables in $\phi$. The meaning of a 2-QBF is that there exists an assignment for the Boolean variables in $V_1$ such that for all assignments to the Boolean variables in $V_2$ the formula $\phi$ over the assigned Boolean variables is satisfied.

For bounded synthesis of Petri games, the bound $b$ is used to build a bounded unfolding $\mathcal{P}^b$ of the Petri game $\mathcal{P}$. $\mathcal{P}^b$ is again a Petri game explicitly modeling all available decision points for the bounded strategy. For a Petri game, the existence of a winning strategy for a play of length $n$ can be encoded as a 2-QBF $\exists\, S \ldotp \forall\, M \ldotp \phi_n$. The set~$S$ describes the strategy and contains pairs $(p,t)$ to indicate whether the system place $p \in \mathcal{P}^b_S$ decides to fire the transition $t \in p^\bullet$ or not. We further ensure that a decision for or against $t$ does not violate justified refusal such that all bounded strategies fulfill this condition. The set~$M$ describing the sequence of markings contains pairs $(p,i)$ to indicate that on place $p$ resides a token at time point $1 \leq i \leq n$. The formula $\phi_n$ ensures that if $M$ represents a play following the decisions by the strategy $S$ and the rules of a Petri game for $n$ steps, then the play is winning. This approach can handle finite and infinite plays by accepting termination before the last simulation step is reached and checking for loops if the last simulation step is reached. 

The encoding for bounded synthesis has the following form:

\begin{eqnarray*}
	\phi_n & \defEQ & \left( \bigwedge_{i\in\{1,\ldots,n-1\}}\mathit{sequence}_i \implies \mathit{win}_i\right) \wedge \left( \mathit{sequence}_n \implies \mathit{win}_n \wedge \mathit{loop} \right)\\
	\mathit{sequence}_i & \defEQ & \mathit{initial} \wedge \bigwedge_{j\in\{1,\ldots,i-1\}}\mathit{flow}_j\\
	\mathit{win}_i & \defEQ &  \mathit{nobadplaces}_i \wedge \mathit{deterministic}_i \wedge \mathit{deadlocksterm}_i\\
	\mathit{deadlocksterm}_i & \defEQ &  \mathit{deadlock}_i \implies \mathit{terminating}_i\\
	\mathit{loop} & \defEQ & \bigvee_{j,k \in\{1,...,n\}, j < k} \left(\bigwedge_{p\in\mathcal{P}} (p,j) \iff (p,k) \right)
\end{eqnarray*}

For each time point $1\leq i \leq n$, it is tested whether the variables in $M$ represent a correct $\mathit{sequence}$ of markings corresponding to a play in the Petri game. If this is the case then \(\mathit{win}_i\) ensures that the strategy fulfills the requirements at $i$ to be winning. If $i=n$, i.e., the limit on the simulation length is reached, then it is additionally tested that a $\mathit{loop}$ occurred. A correct sequence is defined by the play starting from the \emph{initial} marking and firing one enabled and (by the strategy) chosen transition at each time step ($\mathit{flow}_j$). The play is winning if \emph{no bad places} are reached, the system makes only $\mathit{deterministic}$ decisions, and each $\mathit{deadlock}$ is caused by $\mathit{termination}$. A deadlock occurs when all transitions are either not enabled or not chosen by the strategy. Meanwhile, termination occurs when no transition is enabled, i.e., only the lack of tokens in the preset of transitions is responsible for this and not the decisions of the system. Therefore, $\mathit{deadlocksterm}_i$ ensures that the system does not prevent the reaching of bad places by just stopping to fire transitions but deadlocks are only allowed when the whole game terminated. A loop in a Petri game occurs when the exact marking is repeated at two different time points $j$ and $k$. Since it is tested that between $j$ and $k$ the strategy is deterministic, this behavior is repeated infinitely often such that the strategy is also winning in an infinite play. For further details on the definition of $\mathit{initial}$, $\mathit{flow}_j$, $\mathit{deterministic}_i$ etc., we refer the interested reader to~\cite{DBLP:conf/birthday/Finkbeiner15}.

\begin{figure}
 \centering
 \input{./figures/securitySystemBoundedUnfolding}
 \caption{\em A bounded unfolding for the Petri game depicted in Fig.~\ref{fig:SecSystem} modeling an alarm system. The places $\mathit{pA}$ and $\mathit{pB}$ are unfolded three times, respectively, into the places $\mathit{pA}'$, $\mathit{pA}_A$, $\mathit{pA}_B$ and $\mathit{pB}'$, $\mathit{pB}_B$, $\mathit{pB}_A$ whereas the places $\mathit{AA}$, $\mathit{AB}$, $\mathit{BA}$, and $\mathit{BB}$ are not unfolded.}
  \label{fig:SecSystemBoundedUnfolding}
\end{figure}
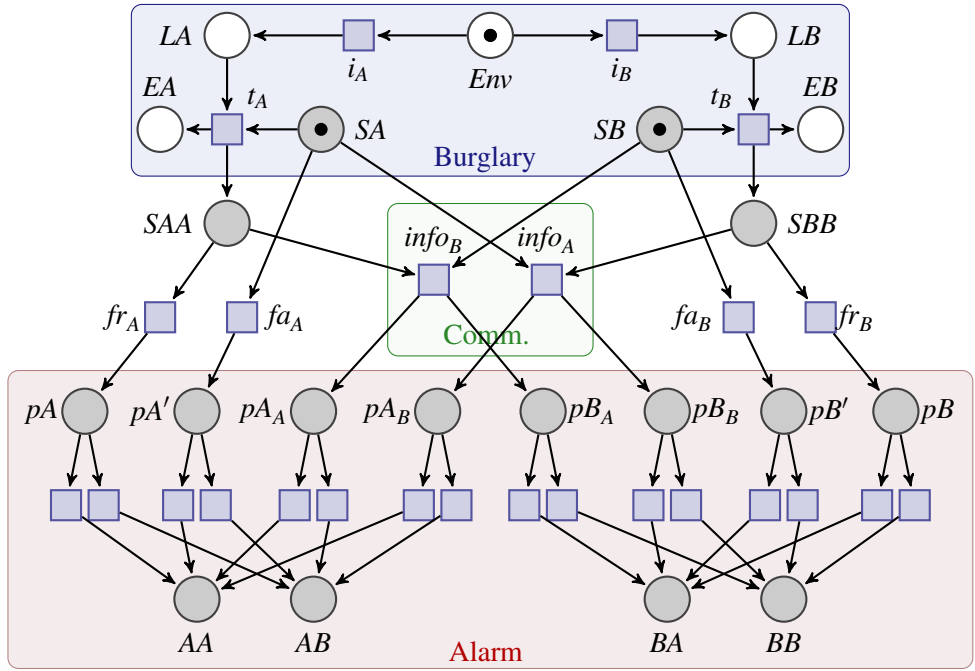

\noindent
{\em
{\bf{}Example.} In Fig.~\ref{fig:SecSystemBoundedUnfolding}, a bounded unfolding of the distributed alarm system is given. Only the places $\mathit{pA}$ and $\mathit{pB}$ are unfolded three times resulting in the four respective places $\mathit{pA}'$, $\mathit{pA}_A$, $\mathit{pA}_B$ and $\mathit{pB}'$, $\mathit{pB}_B$, $\mathit{pB}_A$. The unfolded places of $\mathit{pA}$ are on the lefthand side whereas the unfolded places of $\mathit{pB}$ are on the righthand side. For the four places of $\mathit{pA}$ from left to right, we thereby can differentiate the situation that the alarm system in the corresponding wing successfully tested for the environment and then decided not to inform the other system player (place \(\mathit{pA}\)), did not test for the environment at all (place \(\mathit{pA'}\)), successfully tested for the environment and then informed the other system player (place \(\mathit{pA_A}\)), or was informed by the other system player about an intrusion at the other location (place \(\mathit{pA_B}\)). The same holds in converse direction for $\mathit{pB}$. We did not unfold the places $\mathit{AA}$, $\mathit{AB}$, $\mathit{BA}$, and $\mathit{BB}$ because they are used only to determine bad behavior. This is only possible in the \emph{bounded} unfolding.

We argue why bounded synthesis rejects and accepts certain strategies for the bounded unfolding from Fig.~\ref{fig:SecSystemBoundedUnfolding}. The Petri game is the running example of an alarm system from Fig.~\ref{fig:SecSystem} in Sec.~\ref{sec:recap}. For example, the strategy $(\mathit{SA},t_A)$, $(\mathit{SA},\mathit{fa}_A)$, $(\mathit{SA},\mathit{info}_A)$ activated and all other transitions deactivated is not winning because for the allowed sequence of markings $(\mathit{Env},1)$, $(\mathit{SA},1)$, $(\mathit{SB},1)$, $(\mathit{LA},2)$, $(\mathit{SA},2)$, $(\mathit{SB},2)$ in $M$ there is nondeterminism between the enabled transitions $t_A$ and $\mathit{fa}_A$. Meanwhile, the strategy allowing $(\mathit{SA},t_A)$, $(\mathit{SA}$, $\mathit{info}_A)$, $(\mathit{SB},t_B)$, $(\mathit{SB},\mathit{info}_B)$, $(\mathit{SAA},\mathit{info}_B)$, $(\mathit{SBB},\mathit{info}_A)$, $(\mathit{pA}_A,a)$, $(\mathit{pA}_B,b)$, $(\mathit{pB}_A,a)$, $(\mathit{pB}_B,b)$ is winning because for all markings that represent a valid play of the game no bad place is reached, all decisions are deterministic, and all deadlocks are caused by termination. The places $\mathit{pA}_A$, $\mathit{pA}_B$ and $\mathit{pB}_A$, $\mathit{pB}_B$ describe the unfolded places of $\mathit{pA}$ and $\mathit{pB}$ reached after firing $\mathit{info}_B$ and $\mathit{info}_A$, respectively. For the pairs of unlabeled transitions, the left transitions are based on the original transition $a$ and the right ones on $b$. The places $\mathit{pA}$, $\mathit{pA}'$, $\mathit{pB}$ and $\mathit{pB}'$ are not reached by the strategy and arbitrary decisions can be made there. We choose not to fire any transitions in this case. For example, the sequence of markings $(\mathit{Env},1),(\mathit{SA},1),(\mathit{SB},1),(\mathit{LA},2),(\mathit{LB},2),\dots$ in $M$ does not represent a valid play because both outgoing transitions of $\mathit{Env}$ have been fired, which is illegal in a Petri game ($\mathit{flow}_1$ is violated).
}

\section{Experimental Results}
\label{sec:expResults}

We compare the implementation of the symbolic approach in the tool \adam{} against our prototype implementation of the bounded synthesis approach on an extended set of benchmarks. We take the benchmark set of \adam{} and add the benchmark family of an alarm system. At first, we describe all benchmark families. Then, we outline the technical details of our comparison framework and implementation specific particularities of the two approaches. Afterwards, we state our observations and explanations concerning the times for finding winning strategies and the sizes of these strategies.


\subsection{Benchmark families}

The results in the table from Fig.~\ref{tab:comparison} refer to the following scalable benchmark families where the alarm system is the new benchmark family:
\begin{itemize}
	\item AS: \emph{Alarm System}. There are $m$ secured locations belonging to one person. A burglar can intrude one of the locations. Each location has a local alarm system, which can communicate with all other local alarm systems. The goal is that the alarm system in each location has to indicate the position where the burglar intruded. Furthermore, the alarm system should not issue unsubstantiated warnings of an intrusion at any location.\\\emph{Parameters}: $m$ locations
	\item CM: \emph{Concurrent Machines}. There are $m$ machines which should process $k$ orders. Each machine is allowed to process at most one order. The hostile environment disables one arbitrary machine such that it cannot process any order. The system's goal is to still process all $k$ orders.\\\emph{Parameters}: $m$ machines and $k$ orders.
	\item SR: \emph{Self-reconfiguring Robots}. There are $m$ robots having $m$ different tools at their disposal each. The robots can only equip one tool at a time. From a global perspective, all robots together have to maintain a functioning state such that material can be processed by each of the $m$ different tools. The environment can destroy $k$ tools in total. This can occur at the same robot or on different robots. The robots have to reconfigure theirselves to maintain a functioning global state for the processing of material.\\\emph{Parameters}: $m$ robots with $m$ tools each and $k$ destroyed tools in total.
	\item JP: \emph{Job Processing}. A job requires processing by a, from the environment chosen, subset of $m$ processors in ascending order.\\\emph{Parameter}: $m$ processors.
	\item DW: \emph{Document Workflow}. There are $m$ clerks having to unanimously endorse or reject a document. The environment decides which clerk gets the document first. In DWs, it is required that all clerks endorse the document.\\\emph{Parameter}: $m$ clerks.
\end{itemize}

\subsection{Comparison framework}

We compare the symbolic synthesis approach implemented in \adam{} with our prototype implementation of the bounded synthesis approach. \adam{} and the bounded synthesis approach are the only tools existing to find winning strategies of Petri games but they are inherently different. On the one hand, the symbolic approach models, in theory, infinite memory and unbounded firing sequences of transitions. On the other hand, bounded synthesis has two parameters $n$ and $b$, which can be increased to find a winning strategy. Therefore, the bounded synthesis approach can be parallelized easily because the QBF-solver can be called twice for two different pairs $(n,b)$ and the resulting encodings. We report in the following on the runtime results for the minimal~$b$ and the corresponding~$n$ such that a winning strategy exists because~$b$ turned out to be more expensive than~$n$ in terms of runtime. Notice that $b = 1$ enforces that the bounded unfolding is the original game, i.e., no bounded unfolding is utilized when searching for a winning bounded strategy of the corresponding Petri game.

\begin{figure}
	\centering
	\input{./figures/benchmarksMerged}
	\caption{Comparison between the symbolic synthesis approach and the bounded synthesis approach.}
	\label{tab:comparison}
\end{figure}

The table in Fig.~\ref{tab:comparison} shows the results of \adam{} and our prototype implementation of bounded synthesis for the previously described benchmark families. The results were obtained on an Intel i7-2700K CPU with 3.50~GHz, 32~GB RAM, and a timeout of 1800~seconds. For each benchmark (column \textit{Ben.}), we report on the attempted parameters of the benchmark (\textit{Par.}), on the size of the Petri game (number of tokens ($\#\mathit{Tok}$), places ($\#\pl$), and transitions ($\#\tr$)), and on the respective \emph{time} and \emph{memory} for solving and on the respective number of places ($\#\pl_{\mathit{str}}$) and transitions ($\#\tr_\mathit{str}$) of the winning strategies synthesized by \adam{} and by our prototype implementation of bounded synthesis. The elapsed CPU time is measured in seconds and the used memory in gigabyte. 
For bounded synthesis, we additionally report the smallest~$b$ and corresponding~$n$ to find winning bounded strategies with the least memory requirement. When \adam{} and our prototype implementation both synthesized a winning strategy, then we mark the minimal running time, the minimal memory usage, and the minimal number of places and of transitions in the winning strategy in bold, respectively.

\begin{figure}[h]
    \subfigure[The CPU running time (in minutes) for a selection of benchmarks and the respective number of processes (tokens). The running time for the bounded approach is given as squares and the running time for the symbolic case is given by stars. The dotted lines designate the expected running time after the timeout of 30 minutes.\label{fig:BVa}]{\includegraphics[width=0.49\textwidth]{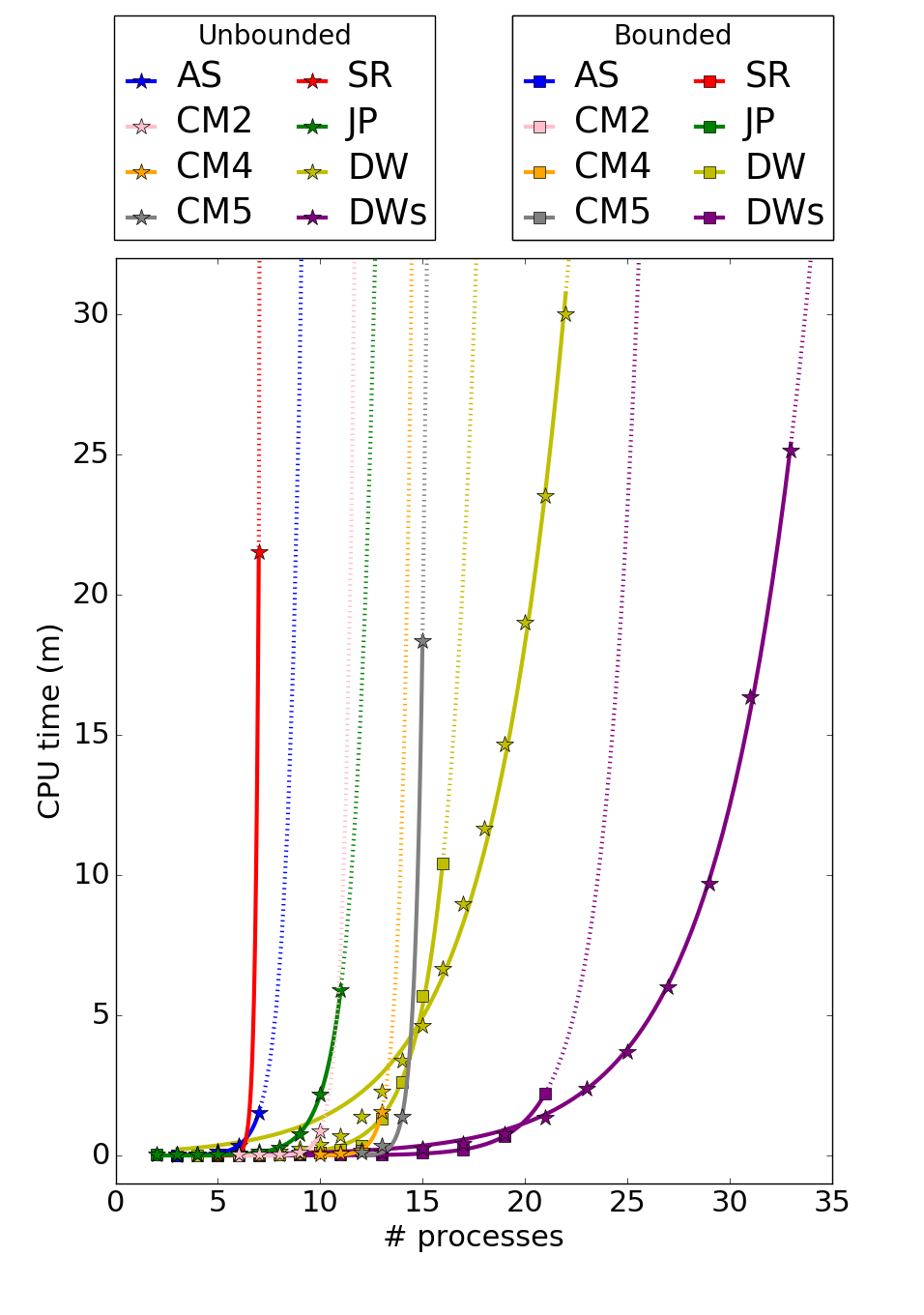}}
\hspace{2mm}%
    \subfigure[The sizes (in the number of transitions) for a selection of benchmarks and the respective number of processes (tokens). The number of transitions of the Petri game is designated by stars, the number of transitions for the bounded strategy by filled circles, and the number of transitions of the strategy in the bounded case by squares.\label{fig:BVb}]{\includegraphics[width=0.49\textwidth]{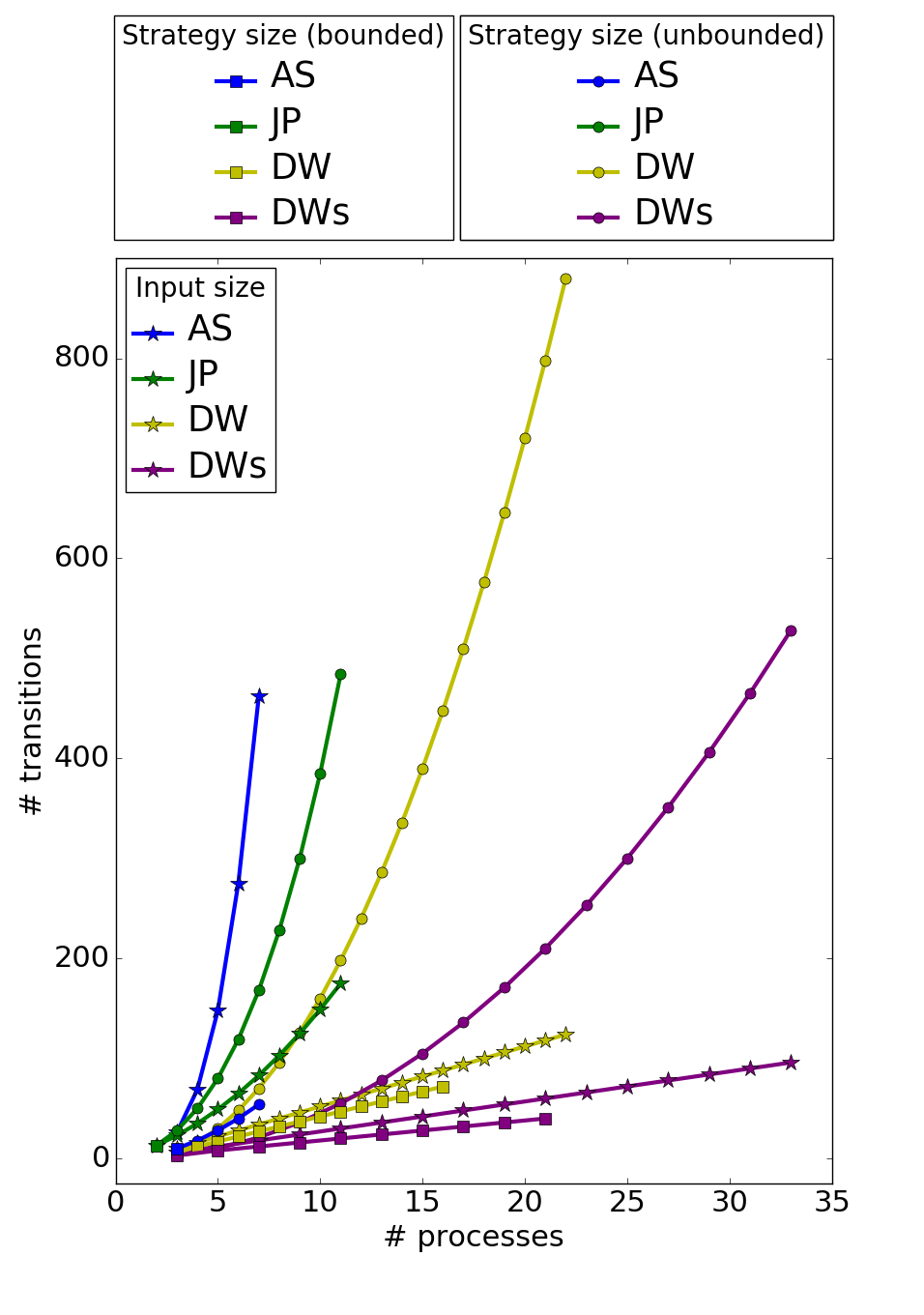}}
\caption{Comparison of the symbolic synthesis approach and the bounded synthesis approach by the running times and sizes of the strategies.}
\label{fig:BV}
\end{figure}

A selection of these values and benchmark families are plotted in Fig.~\ref{fig:BV}. In Fig.~\ref{fig:BVa}, the CPU times in minutes for the symbolic and the bounded approach on selected benchmark families are plotted for an increasing number of processes, i.e., the number of tokens of the underlying net. \(CMi\), for \(i\in\{2,4,5\}\), represents the subset of the concurrent machines benchmark, where the first parameter~$m$ for the number of machines is fixed to~\(i\). Therefore, the number of orders increases the number of processes in \(CMi\). The dotted lines designate the expected running time obtained by fitting exponential curves through the actual values. In Fig.~\ref{fig:BVb}, we plotted the number of transitions of the input Petri games and of the corresponding winning strategies of the two approaches for an increasing number of processes from selected benchmark families.

\subsection{Implementation details}
\label{sec:impl}
During the implementation of our prototype for bounded synthesis of Petri games, we observed that a translation of the matrix $\phi_n$ into conjunctive normalform~(CNF) and the usage of a QBF solver requiring input in CNF has poor performance. We found out that the \texttt{QCIR} file format for QBFs~\cite{qcir} allows competitive performance as $\phi_n$ does not need to be translated into CNF. As we have a 2-QBF not in CNF, solvers using counterexamples to refine an abstraction~(CEGAR-based)~\cite{DBLP:conf/cav/ClarkeGJLV00} showed the best performance. We therefore decided for the solver \textsc{QuAbS}~\cite{conf/sat/Tentrup16} as it combines fast parsing with fast solving. Given a bound on the length of the proof of correctness for a strategy, we further pruned the bounded unfolding of unreachable places and unreachable transitions to remove unnecessary variables from the 2-QBF, which increased the overall performance.

The running times of both approaches are highly dependent on the number of variables which are respectively used in the BBD and in the QBF. 
For the symbolic approach, the number of variables in the BDD grows significantly with the number of players in the Petri game (which are represented by tokens in the underlying net).
The two-player game over a finite graph with complete information is represented by a BDD for each flow encoding the flow's source and target state.
For optimizing the size of the BDD, we partition \(\mathcal{P}_S\) into \(k\) disjunct sets \(\mathcal{P}_{S_i}\subseteq\mathcal{P}_S\), for \(i\in\{1,\ldots,k\}\),
such that for every reachable marking each place of the marking belongs uniquely to exactly one of the sets \(\mathcal{P}_{S_i}\). 
In general, the number \(k\) corresponds to the number of system processes in the game. A state of the finite graph is encoded by a binary encoding of the ID of the environment place and the maximally \(k\) IDs of the system places of the marking, which results in encodings of logarithmic size. Whether a process's commitment set contains a transition is encoded explicitly by a Boolean flag. An explicit encoding is used because the size of the commitment set varies for each process depending on the place it resides in. In general, this approach yields smaller BDDs. The two additional variables per system process encode whether the process's commitment set has to be renewed, i.e., the \(\top\)-flag is set or not, and whether the place is a $\type_2$-typed place. Thus, the number of variables for a BDD can be calculated by
\[2\cdot \left(\mathtt{log_2}(\,|\mathcal{P}_E|\,) + \sum_{i=1}^{k}\left(\mathtt{log_2}(\,|\mathcal{P}_{{S}_i}|\,)+\,|\tr_i|\,+2\right)\right)\]
where $\tr_i=\bigcup_{p\in{\pl_{S}}_i}\postset{p}$.


For the bounded approach, we distinguish existentially and universally quantified variables and gate variables.
The existentially quantified variables describe the system's strategy and the universally quantified variables encode all sequences of markings.
The gate variables are used to describe the bounded synthesis problem in the matrix $\phi_n$ given the existentially and universally quantified variables.
Thus, the size of the QBF is growing with the structure of the underlying net of the Petri game, i.e., with the number of places and transitions.
	
\begin{figure}
	\centering
	\input{./figures/tableVariableSize}
	\caption{Comparison between the numbers of different variables in the two approaches.}
	\label{tab:variablesize}
\end{figure}

In Fig.~\ref{tab:variablesize}, we compare the number of variables in the two approaches for benchmark families where several instances are solved by both approaches. DW, DWs, and CM with parameter $k=1$ qualify for this comparison. For the symbolic approach, the number of variables in the BDD is given ($\#\mathit{Var}_\mathit{symbolic}$). For the bounded approach, the total number of variables in the QBF ($\#\mathit{Var}_\mathit{bounded}$) and the number of existentially ($\#\mathit{Var}_\exists$) and universally ($\#\mathit{Var}_\forall$) quantified variables thereof are given. The number of gate variables ($\#\mathit{Var}_{\phi_n}$) used to build the matrix $\phi_n$ is stated as the remaining variables of $\#\mathit{Var}_\mathit{bounded}$, i.e., $\#\mathit{Var}_\exists + \#\mathit{Var}_\forall + \#\mathit{Var}_{\phi_n} = \#\mathit{Var}_\mathit{bounded}$. From the size difference of the respective numbers of variables in the two approaches, one can derive that they are used for a different purpose in the respective approach. For the symbolic approach, the number of variables grows linearly for all benchmarks. For the bounded approach, the number of existentially quantified variables grows linearly, the number of universally quantified variables grows quadratically, and the total number of variables grows cubicly in DW and DWs. For CM, all variables in the bounded encoding grow exponentially which is surprising as $n$ and $b$ remain constant. We suppose that this increase is caused by the construction of the bounded unfolding which produces an exponentially growing number of transitions for this benchmark family despite $b=3$ remaining constant.

We further detected that the QBF problem files can become large, which requires a QBF solver with fast parsing of the problem file. The largest solved \texttt{QCIR} file is of size 40~MB and contains $208.877$ variables (benchmark \emph{concurrent machines} with parameters $m=15$ and $k=1$ and bounds $n=6$ and $b=3$). A very large \texttt{QCIR} file for a benchmark which \adam{} solved but for which the QBF solver timed out (benchmark \emph{job processing} with parameter $m=4$ and bounds $n=9$ and $b=5$) has a size of 275~MB and contains $2.722.512$ variables. 

In summary, the size of the BDD for the symbolic approach is dominated by the number of tokens whereas the size of the 2-QBF for the bounded approach is dominated by the number of places and transitions of the underlying net of the Petri game.

\subsection{Comparison}

Both approaches work especially well on certain aspects of distributed synthesis. The symbolic approach implemented in \adam{} solves more instances than the bounded synthesis approach for all benchmark families but for \emph{concurrent machines}~(CM) with one defective machine ($k=1$). \adam{} can further show the non-existence of a winning strategy for instantiations of benchmark families for which bounded synthesis is not applicable. For example, \adam{} shows that for CM no strategy exists when equally many or more orders as machines are placed because one machine can process at most one order and the environment marks one machine as unable to process an order.

The bounded approach is well suited for finding small winning strategies. It holds for all winning strategies produced by the two approaches that the respective winning strategies from bounded synthesis have equally many or fewer places and transitions. For small instances, these differences are negligible as for the first instances of \emph{alarm system} (AS) and both versions of \emph{document workflow} (DW and DWs) the respective strategies are of equal size. The larger the benchmark instances become, the larger the differences in strategy size get. For~DW with parameter $m=14$, the strategy from \adam{} has 660 places and 448 transitions whereas the strategy from the prototype implementation of bounded synthesis has only 88 places and 72 transition. This comes at a  higher solving time of 625 seconds in contrast to 400~seconds and at using 12~GB of memory in comparison to 4~GB. 

The difference in size of the strategies can also be observed from Fig.~\ref{fig:BVb} where the size of the input and of the produced strategies by the two approaches are compared for a given number of processes. This difference in size is based on the different structure of the strategies in the two approaches. In bounded unfoldings and bounded strategies, more than one transition can merge into one place, if the different history of the token is not needed. In contrast, the symbolic approach has to unfold a place in every case notwithstanding the need for differentiation of its causal past. This becomes apparent in the benchmarks DW and DWs. In the symbolic case, for every choice of the environment which clerk has to decide on endorsing the document first, the places and transitions of each clerk are copied and put into the right order. The bounded algorithm detects that it is not necessary to unfold all these places, since the memory is not needed for finding a winning strategy and thus yields a much smaller strategy. 

The bounded approach can be more subtle in choosing when to unfold places and therewith generally finds smaller strategies than the symbolic approach. This illustrates the difference between a bounded strategy (produced by the bounded synthesis approach) and a strategy (produced by the symbolic synthesis approach). Bounded strategies are based on the bounded unfolding which can consolidate different causal pasts into one system place for which the corresponding strategy has to make the same decision. In contrast, a strategy is based on the unfolding, which explicitly represents every causal past of a system place. At each such system place, an individual decision can be made. Therefore, when the same decision suffices for each causal past, these decisions are represented explicitly with each unfolded place. From the visualization of Fig.~\ref{fig:BVb}, we can conjecture for the displayed subset of benchmark families that the symbolic approach can only find strategies which grow in size exponentially because the unfolding is exponential in the number of places and transitions. In contrast, the bounded approach can find strategies which grow in size linearly when a linearly growing bounded unfolding suffices to represent the necessary causal history.

For the sizes of the solution in the symbolic case, we can see that in general the strategy sizes increase faster and also are larger than the sizes of the input
in the benchmark families from Fig.~\ref{fig:BVb}. This is caused by our benchmarks mostly increasing linearly in the input sizes (e.g., by adding a new robot
or a new machine). Meanwhile, the solution is getting more difficult due to the additional abilities and behaviors
of the whole system (e.g., the factory) and the symbolic approach has to consider more different flows of tokens with different causal histories.
In general, the unfolding and the strategy increase in size stronger than the input. One exception to the linear increase of the input is the new \emph{alarm system} benchmark. There we also add only linearly bounded many places 
and transitions for every new alarm system, but we have to add an additional alarm signal at each already existing alarm system and, furthermore, transitions
leading to bad places for all additional combinations of bad situations. Since those transitions are not present in the strategy, the size of the
solution is smaller than the size of the input for the alarm system benchmark.

For the running time, we can see in Fig.~\ref{tab:comparison} and in Fig.~\ref{fig:BVa} that the bounded approach outperforms the symbolic one for smaller instances but increases more sharply.
This stems from the different parameters in both approaches discussed in Sec.~\ref{sec:impl}, which are responsible for the solving complexity.
For the symbolic case, the number
of processes are principally responsible for the increasing number of variables of the BDD and thus for the complexity.
For the bounded approach, this is different because the number of QBF variables is more dependent on 
the structure of the net than on the number of tokens.

For the benchmarks of DW and DWs, the bounded synthesis approach outperforms \adam{} for the first eleven respectively nine parameters in terms of runtime and memory usage. On the next three parameters (DW) respectively on the next parameter (DW), \adam{} outperforms bounded synthesis. After that, bounded synthesis already reaches the time limit while \adam{} can solve further five parameters for DW and DWs each. 

The bounded synthesis approach further showed that no unfolding is necessary to solve instances of~DW and~DWs. It also revealed that for CM it is possible to find winning strategies for benchmark instances of growing size while maintaining the same values for $n$ and $b$.

\section{Conclusion}
\label{sec:conlusion}

We added the new benchmark family of a distributed alarm system to the set of benchmark families for distributed synthesis with Petri games collected during the implementation of \adam{}. The new benchmark family models an alarm system for a person with a scalable number of independent locations she needs to secure. Each of these locations has a local alarm system, which can detect the intrusion by a burglar. Furthermore, all alarm systems can communicate with each other and each alarm system can indicate the position of a detected burglary. We synthesized strategies to detect the position of a burglary and indicate it at the alarm systems of \emph{all} independent locations.

We found out that the translation of bounded synthesis into 2-QBF resulted in large non-CNF formulas, which were solved best by a CEGAR-based QBF solver like \textsc{QuAbS}. The automatic construction of the bounded unfolding benefits from a removal of unreachable places and transitions, which implies that there is still room for improvement when constructing the bounded unfolding.

We compared the symbolic synthesis approach implemented in the tool \adam{} with the bounded synthesis approach on the extended set of benchmarks. We found out that symbolic synthesis can overall synthesize strategies for larger problems for all benchmark families except the benchmark family of concurrent machines (with parameter $k=1$). For the smaller instances, bounded synthesis is faster but the running time grows at a higher rate such that \adam{} can solve more instances overall. At the same time, bounded synthesis finds smaller strategies in the number of places and transitions. This difference is negligible for small instances but grows for larger instances. The difference in size of the strategies is caused by the distinction between the bounded unfolding and the unfolding. In the bounded unfolding, different causal pasts can be consolidated into one place whereas in the unfolding, unique causal pasts have to be differentiated. Therefore, bounded strategies can profit in terms of size from situations where only some causal past is needed. This adds to the general benefit of bounded synthesis in comparison with symbolic synthesis to steer the search to small strategies.

We identified that the number of variables in the BDD of the symbolic approach grows in the number of tokens in the underlying net of the Petri game whereas the number of variables in the QBF of the bounded approach grows in the number of places and transitions of the underlying net of the Petri game. We further showed that for the benchmark families DW and DWs local strategies for each system player suffice as the bounded unfolding is the same as the original game. This proved that both symbolic and bounded synthesis are well-suited for certain aspects of distributed synthesis with Petri games.


\nocite{*}
\bibliographystyle{eptcs}
\bibliography{bibliography}
\end{document}

%% file: figures/securitySystem.tex
\begin{tikzpicture}[node distance=1.25cm,>=stealth',bend angle=45,auto,scale=0.8]
	\node [envplace,tokens=1] (env)  [label=below:$\mathit{Env}$, tokens=1]                                  {};
	\node [transition] (eL)  [left of=env, xshift=-5mm,label={[label distance=-0.5mm]below:$i_A$}] {};
	\node [transition] (eR)  [right of=env, xshift=5mm,label={[label distance=-0.5mm]below:$i_B$}] {};
	\node [envplace] (epL)  [label=left:$\mathit{LA}$, left of=eL, xshift=-5mm]                                  {};
	\node [envplace] (epR)  [label=right:$\mathit{LB}$, right of=eR, xshift=5mm]                                  {};
	\node [transition] (tA)  [below of=epL,label={[label distance=-1mm]above right:$t_A$}] {};
	\node [sysplace] (IA) [below of=tA,label=left:$\mathit{SAA}$]                     {};
	\node [envplace] (ea)  [left of=tA,xshift=3.6mm,label=above:$\mathit{EA}$]                                  {};
	\node [sysplace] (A) [right of=tA,label=right:$\mathit{SA}$,tokens=1]                     {};
	\node [transition] (t2)  [below of=IA,xshift=2mm,label=right:\(\mathit{fa}_A\)] {};
	\node [transition] (t1)  [left of=t2,xshift=1.6mm,label={[label distance=-1mm]right:\(\mathit{fr}_A\)}] {};
	\node [transition] (t3)  [right of=t2,xshift=13mm,yshift=5mm,label=above:$\mathit{info}_B$] {};
	\node [sysplace] (pa) [below of=t2,xshift=-2mm,label=left:$\mathit{pA}$]                     {};
	\node [transition] (tAB)  [below left of=pa,label=right:$\mathit{aa}\phantom{b}$] {};
	\node [transition] (tAA)  [below right of=pa,label=left:$\mathit{ab}$] {};
	\node [sysplace] (ab) [below of=tAB,label=right:$\mathit{AA}$]                     {};
	\node [sysplace] (aa) [below of=tAA,label=right:$\mathit{AB}$]                     {};
	\node [transition] (tB)  [below of=epR,label={[label distance=-1mm]above left:$t_B$}] {};
	\node [sysplace] (IB) [below of=tB,label=right:$\mathit{SBB}$]                     {};
	\node [envplace] (eb)  [right of=tB,xshift=-3.6mm,label=above:$\mathit{EB}$]                                  {};
	\node [sysplace] (B) [left of=tB,label=left:$\mathit{SB}$,tokens=1]                     {};
	\node [transition] (t2b)  [below of=IB,xshift=-2mm,label=left:\(\mathit{fa}_B\)] {};
	\node [transition] (t1b)  [right of=t2b,xshift=-1.6mm,label={[label distance=-1mm]left:\(\mathit{fr}_B\)}] {};
	\node [transition] (t3b)  [left of=t2b,xshift=-13mm, yshift=5mm,label=above:$\mathit{info}_A$] {};
	\node [sysplace] (pb) [below of=t2b,xshift=2mm,label=right:$\mathit{pB}$]                     {};
	\node [transition] (tBB)  [below left of=pb,label=right:$\mathit{ba}$] {};
	\node [transition] (tBA)  [below right of=pb,label=left:$\mathit{bb}$] {};
	\node [sysplace] (bb) [below of=tBB,label=left:$\mathit{BA}$]                     {};
	\node [sysplace] (ba) [below of=tBA,label=left:$\mathit{BB}$]                     {};

\node [right=of A,  xshift=-2mm, yshift=-4mm] (intr) {\DarkBlue{}Burglary};
\node [below=of t3, xshift=7mm,  yshift=10mm] (com) {\color{ganttGreen}Comm.};
\node [right=of ab, xshift=22mm, yshift=-2mm] (al) {\Red{}Alarm};

	\path[-latex, thick] 	(eL)  		edge [pre]                            (env)
					edge [post]                            (epL)
		 	(tA)  		edge [pre]                            (epL)
		 	  		edge [pre]                            (A)
					edge [post]                            (ea)
					edge [post]                            (IA)
			(t1)  		edge [pre]                            (IA)
					edge [post]                            (pa)
			(t2)  		edge [pre]                            (A)
					edge [post]                            (pa)
			(t3)  		edge [pre]                            (IA)
			  		edge [pre]                            (B)
					edge [post]                            (pa)
					edge [post]                            (pb)
			(tAB)  		edge [pre]                            (pa)
					edge [post]                            (ab)
			(tAA)  		edge [pre]                            (pa)
					edge [post]                            (aa);
	\path[->, thick] 	(eR)  		edge [pre]                            (env)
					edge [post]                            (epR)
		 	(tB)  		edge [pre]                            (epR)
		 	  		edge [pre]                            (B)
					edge [post]                            (eb)
					edge [post]                            (IB)
			(t1b)  		edge [pre]                            (IB)
					edge [post]                            (pb)
			(t2b)  		edge [pre]                            (B)
					edge [post]                            (pb)
			(t3b)  		edge [pre]                            (IB)
			  		edge [pre]                            (A)
					edge [post]                            (pb)
					edge [post]                            (pa)
			(tBB)  		edge [pre]                            (pb)
					edge [post]                            (bb)
			(tBA)  		edge [pre]                            (pb)
					edge [post]                            (ba);

\begin{pgfonlayer}{background}
\draw [-, rectangle,rounded corners,DarkBlue,fill=cdcBlueL!15] ([xshift=-2mm,yshift=18mm]ea.north west) -- ([xshift=2mm,yshift=18mm]eb.north east) -- ([xshift=2mm,yshift=-5mm]eb.south east) -- ([xshift=-2mm,yshift=-5mm]ea.south west) -- cycle;
\draw [-, rectangle,rounded corners,DarkRed,fill=DarkRed!15, fill opacity=0.95] ([xshift=-2mm,yshift=-3mm]ab.south west) -- ([xshift=-2mm,yshift=35mm]ab.south west) -- ([xshift=2mm,yshift=35mm]ba.south east) -- ([xshift=2mm,yshift=-3mm]ba.south east) -- cycle;
\draw [-, rectangle,rounded corners,ganttGreen,fill=cdcGreenL!15, fill opacity=0.45] ([xshift=-5mm,yshift=-10mm]t3.south west) -- ([xshift=-5mm,yshift=10mm]t3.north west) -- ([xshift=5mm,yshift=10mm]t3b.north east) -- ([xshift=5mm,yshift=-10mm]t3b.south east) -- cycle;
\end{pgfonlayer}
\end{tikzpicture}

%% file: figures/securitySystemStrat.tex
\begin{tikzpicture}[node distance=1.25cm,>=stealth',bend angle=45,auto,scale=0.8]
	\node [envplace,tokens=1] (env)  [label=below:$\mathit{Env}$, tokens=1]                                  {};
	\node [transition] (eL)  [left of=env, xshift=-5mm,label={[label distance=-0.5mm]below:$i_A$}] {};
	\node [transition] (eR)  [right of=env, xshift=5mm,label={[label distance=-0.5mm]below:$i_B$}] {};
	\node [envplace] (epL)  [label=left:$\mathit{LA}$, left of=eL, xshift=-5mm]                                  {};
	\node [envplace] (epR)  [label=right:$\mathit{LB}$, right of=eR, xshift=5mm]                                  {};
	\node [transition] (tA)  [below of=epL,label={[label distance=-1mm]above right:$t_A$}] {};
	\node [sysplace] (IA) [below of=tA,label=left:$\mathit{SAA}$]                     {};
	\node [envplace] (ea)  [left of=tA,xshift=3.6mm,label=above:$\mathit{EA}$]                                  {};
	\node [sysplace] (A) [right of=tA,label=right:$\mathit{SA}$,tokens=1]                     {};
	\node [transition] (t3)  [right of=t2,xshift=13mm,yshift=5mm,label=above:$\mathit{info}_B$] {};
	\node [sysplace] (pa) [below of=t2,xshift=-10.9mm,label=right:$\mathit{pA}$]                     {};
	\node [sysplace] (pa2) [below of=t2,xshift=+2mm,label=right:$\mathit{pA'}$]                     {};
	\node [transition] (tAB)  [below of=pa,label=right:$\mathit{aa\phantom{b}}$] {};
	\node [transition] (tAA)  [below of=pa2,label=right:$\mathit{ab}$] {};
	\node [sysplace] (ab) [below of=tAB,label=right:$\mathit{AA}$]                     {};
	\node [sysplace] (aa) [below of=tAA,label=right:$\mathit{AB}$]                     {};
	\node [transition] (tB)  [below of=epR,label={[label distance=-1mm]above left:$t_B$}] {};
	\node [sysplace] (IB) [below of=tB,label=right:$\mathit{SBB}$]                     {};
	\node [envplace] (eb)  [right of=tB,xshift=-3.6mm,label=above:$\mathit{EB}$]                                  {};
	\node [sysplace] (B) [left of=tB,label=left:$\mathit{SB}$,tokens=1]                     {};
	\node [transition] (t3b)  [left of=t2b,xshift=-13mm, yshift=5mm,label=above:$\mathit{info}_A$] {};
	\node [sysplace] (pb) [below of=t2b,xshift=10.9mm,label=left:$\mathit{pB}$]                     {};
	\node [sysplace] (pb2) [below of=t2b,xshift=-2mm,label=left:$\mathit{pB'}$]                     {};
	\node [transition] (tBB)  [below of=pb2,label=left:$\mathit{ba}$] {};
	\node [transition] (tBA)  [below of=pb,label=left:$\mathit{bb}$] {};
	\node [sysplace] (bb) [below of=tBB,label=left:$\mathit{BA}$]                     {};
	\node [sysplace] (ba) [below of=tBA,label=left:$\mathit{BB}$]                     {};

\node [right=of A, yshift=-4mm, xshift=-2mm] (intr) {\DarkBlue{}Burglary};
\node [below=of t3, xshift=7mm, yshift=10mm] (com) {\color{ganttGreen}Comm.};
\node [right=of ab, xshift=22mm, yshift=-2mm] (al) {\Red{}Alarm};

	\path[-latex, thick] 	(eL)  		edge [pre]                            (env)
					edge [post]                            (epL)
		 	(tA)  		edge [pre]                            (epL)
		 	  		edge [pre]                            (A)
					edge [post]                            (ea)
					edge [post]                            (IA)
			(t3)  		edge [pre]                            (IA)
			  		edge [pre]                            (B)
					edge [post]                            (pa)
					edge [post]                            (pb2)
			(tAB)  		edge [pre]                            (pa)
					edge [post]                            (ab)
			(tAA)  		edge [pre]                            (pa2)
					edge [post]                            (aa);
	\path[->, thick] 	(eR)  		edge [pre]                            (env)
					edge [post]                            (epR)
		 	(tB)  		edge [pre]                            (epR)
		 	  		edge [pre]                            (B)
					edge [post]                            (eb)
					edge [post]                            (IB)
			(t3b)  		edge [pre]                            (IB)
			  		edge [pre]                            (A)
					edge [post]                            (pb)
					edge [post]                            (pa2)
			(tBB)  		edge [pre]                            (pb2)
					edge [post]                            (bb)
			(tBA)  		edge [pre]                            (pb)
					edge [post]                            (ba);

\begin{pgfonlayer}{background}
\draw [-, rectangle,rounded corners,DarkBlue,fill=cdcBlueL!15] ([xshift=-2mm,yshift=18mm]ea.north west) -- ([xshift=2mm,yshift=18mm]eb.north east) -- ([xshift=2mm,yshift=-5mm]eb.south east) -- ([xshift=-2mm,yshift=-5mm]ea.south west) -- cycle;
\draw [-, rectangle,rounded corners,DarkRed,fill=DarkRed!15, fill opacity=0.95] ([xshift=-2mm,yshift=-3mm]ab.south west) -- ([xshift=-2mm,yshift=40mm]ab.south west) -- ([xshift=2mm,yshift=40mm]ba.south east) -- ([xshift=2mm,yshift=-3mm]ba.south east) -- cycle;
\draw [-, rectangle,rounded corners,ganttGreen,fill=cdcGreenL!15, fill opacity=0.45] ([xshift=-5mm,yshift=-10mm]t3.south west) -- ([xshift=-5mm,yshift=10mm]t3.north west) -- ([xshift=5mm,yshift=10mm]t3b.north east) -- ([xshift=5mm,yshift=-10mm]t3b.south east) -- cycle;
\end{pgfonlayer}
\end{tikzpicture}

%% file: figures/securitySystemDistrStrat.tex
\begin{tikzpicture}[node distance=5mm and 10mm,>=stealth',bend angle=45,auto,scale=0.8]
	\node [envplace,tokens=1] (env)  		[label=above:$\mathit{Env}$, tokens=1]                                  {};
	\node [transition,below left=of env] (eL)  	[label={[label distance=-0.5mm]right:$i_A$}] {};
	\node [transition,below right=of env] (eR)	[label={[label distance=-0.5mm]left:$i_B$}] {};
	\node [envplace, below=of eL] (epL)  		[label=right:$\mathit{LA}$]                                  {};
	\node [envplace, below=of eR] (epR)  		[label=left:$\mathit{LB}$]                                  {};
	\node [transition] (tA1) 			[below=of epL,label={[label distance=-0.5mm]right:$t_A$}] {};
	\node [transition] (tB1)  			[below=of epR,label={[label distance=-0.5mm]left:$t_B$}] {};
	\node [envplace] (ea)  				[below=of tA1,xshift=0mm,label=right:$\mathit{EA}$]                                  {};
	\node [envplace] (eb)  				[below=of tB1,xshift=0mm,label=left:$\mathit{EB}$]                                  {};

		\node [sysplace, xshift=45mm] (A) [right=of env,label=above:$\mathit{SA}$,tokens=1]                     {};	
		\node [transition] (tA2)  [below left=of A,label={right:$t_A$}] {};
		\node [transition] (t3b1)  [below right=of A,label=left:$\mathit{info}_A$] {};
		\node [sysplace] (IA) [below=of tA2,label=right:$\mathit{SAA}$]                     {};
		\node [transition] (t3)  [below=of IA,label=right:$\mathit{info}_B$] {};
		\node [sysplace] (pa) [below=of t3,label=right:$\mathit{pA}$]                     {};
		\node [transition] (tAB)  [below=of pa,label=right:$\mathit{aa\phantom{b}}$] {};
		\node [sysplace] (ab) [below=of tAB,label=right:$\mathit{AA}$]                     {};
		\node [sysplace, xshift=12.5mm] (pa2) [right=of pa,label=right:$\mathit{pA'}$]                     {};
		\node [transition] (tAA)  [below=of pa2,label=right:$\mathit{ab}$] {};
		\node [sysplace] (aa) [below=of tAA,label=right:$\mathit{AB}$]                     {};
		\node [sysplace, xshift=45mm] (B) [right=of A,label=above:$\mathit{SB}$,tokens=1]                     {};
		\node [transition] (tB)  [below right=of B,label={left:$t_B$}] {};
		\node [transition] (t31)  [below left=of B, label=right:$\mathit{info}_B$] {};
		\node [sysplace] (IB) [below=of tB,label=left:$\mathit{SBB}$]                     {};
		\node [transition] (t3b2)  [below=of IB,label=left:$\mathit{info}_A$] {};
		\node [sysplace] (pb2) [below=of t3b2,label=left:$\mathit{pB}$]                     {};
		\node [transition] (tBB)  [below=of pb2,label=left:$\mathit{bb}$] {};
		\node [sysplace] (bb) [below=of tBB,label=left:$\mathit{BB}$]                     {};
		\node [sysplace, xshift=-12.5mm] (pb) [left=of pb2,label=right:$\mathit{pB'}$]                     {};
		\node [transition] (tBA)  [below=of pb,label=right:$\mathit{ba}$] {};
		\node [sysplace] (ba) [below=of tBA,label=right:$\mathit{BA}$]                     {};

\node [right=of env, yshift=5mm, xshift=8mm] (intr) {\DarkBlue{}Burglary};
\node [right=of t3b1, xshift=-9.5mm, yshift=0mm] (com) {\color{ganttGreen}Comm.};
\node [right=of aa, xshift=-1mm, yshift=-2mm] (al) {\Red{}Alarm};

	\path[-latex, thick] 	(eL)  		edge [pre]                            (env)
					edge [post]                            (epL)
		 	(tA1)  		edge [pre]                            (epL)
					edge [post]                            (ea)
		 	(eR)  		edge [pre]                            (env)
					edge [post]                            (epR)
		 	(tB1)  		edge [pre]                            (epR)
					edge [post]                            (eb);

\path[-latex, thick] 	(tA2)  		edge [pre]                            (A)					
					edge [post]                            (IA)
			(t3b1) 		edge [pre]                            (A)
					edge [post]                            (pa2)
			(t3)  		edge [pre]                            (IA)
					edge [post]                            (pa)
			(tAB)  		edge [pre]                            (pa)
					edge [post]                            (ab)
			(tAA)  		edge [pre]                            (pa2)
					edge [post]                            (aa)
;

\path[-latex, thick] 	(tB)  		edge [pre]                            (B)					
					edge [post]                            (IB)
			(t31) 		edge [pre]                            (B)
					edge [post]                            (pb)
			(t3b2) 		edge [pre]                            (IB)
					edge [post]                            (pb2)
			(tBA)  		edge [pre]                            (pb)
					edge [post]                            (ba)
			(tBB)  		edge [pre]                            (pb2)
					edge [post]                            (bb)
;

	\node (p1)  [right of=env, xshift=22.5mm, yshift=-5mm]                                  {};
	\node (p2)  [below of=p1,              yshift=-45mm]                                  {};
	\draw[thick]		(p1.east)  -- (p2.east);
	\draw[thick]		([xshift=2mm]p1.east)  -- ([xshift=2mm]p2.east);

	\node (p3)  [right of=A, xshift=25mm, yshift=-5mm]                                  {};
	\node (p4)  [below of=p3,              yshift=-45mm]                                  {};
	\draw[thick]		(p3.east)  -- (p4.east);
	\draw[thick]		([xshift=2mm]p3.east)  -- ([xshift=2mm]p4.east);

\begin{pgfonlayer}{background}
\draw [-, rectangle,rounded corners,DarkBlue,fill=cdcBlueL!15] ([xshift=-2mm,yshift=-2mm]ea.south west) -- ([xshift=5mm,yshift=-2mm]eb.south east) -- ([xshift=5mm,yshift=-2mm]eR.south east) -- ([xshift=2mm,yshift=-2mm]tA2.south east) -- ([xshift=2mm,yshift=-2mm]A.south east) -- ([xshift=-2mm,yshift=-2mm]B.south east) -- ([xshift=-2mm,yshift=-2mm]tB.south west) -- ([xshift=2mm,yshift=-2mm]tB.south east)  -- ([xshift=2mm,yshift=25mm]tB.south east)  -- ([xshift=-196.5mm,yshift=25mm]tB.south east) -- cycle;
\draw [-, rectangle,rounded corners,DarkRed,fill=DarkRed!15, fill opacity=0.95] ([xshift=-2mm,yshift=-3mm]ab.south west) -- ([xshift=-2mm,yshift=36mm]ab.south west) -- ([xshift=2mm,yshift=36mm]bb.south east) -- ([xshift=2mm,yshift=-3mm]bb.south east) -- cycle;
\draw [-, rectangle,rounded corners,ganttGreen,fill=cdcGreenL!15, fill opacity=0.45] ([xshift=-2mm,yshift=-2mm]t3.south west) -- ([xshift=-2mm,yshift=2mm]t3.north west)  -- ([xshift=16mm,yshift=2mm]t3.north west) -- ([xshift=-12mm,yshift=2mm]t3b1.north west) -- ([xshift=12mm,yshift=2mm]t31.north east) -- ([xshift=-10mm,yshift=2mm]t3b2.north west) -- ([xshift=2mm,yshift=2mm]t3b2.north east) --  ([xshift=2mm,yshift=-2mm]t3b2.south east) --  ([xshift=-25mm,yshift=-2mm]t3b2.south west) --  ([xshift=0mm,yshift=-2mm]t31.south)  --  ([xshift=0mm,yshift=-2mm]t3b1.south)  -- ([xshift=25mm,yshift=-2mm]t3.south east)  -- cycle;
\end{pgfonlayer}
\end{tikzpicture}

%% file: figures/securitySystemGG.tex
\begin{tikzpicture}[
sys/.style={
rectangle,
rounded corners,
very thick,
fill=lightgray,
draw,
align=center,
minimum size=7mm,
},
env/.style={
rectangle,
very thick,
fill=white,
draw,
align=center,
minimum size=7mm,
},
node distance=7mm and 2mm
]
\node[sys,win, dashed,label={[yshift=-5mm, xshift=1mm]above left:\circled{1}}]	(init)	{\(\mathit{Env}\),\\\((\mathit{SA}, 1, \top, \{\})\),\\\((\mathit{SB}, 1, \top, \{\})\)};
\node[env,win, below=of init]		(0)	{\(\mathit{Env}\),\\\((\mathit{SA}, 1, !\top, \{t_A,\mathit{info_A}\})\),\\\((\mathit{SB}, 1, !\top, \{t_B,\mathit{info_B}\})\)};

\node[env,win, below=of 0] 	(2)	{\(\mathit{LA}\),\\\((\mathit{SA}, 1, !\top, \{t_A,\mathit{info_A}\})\),\\\((\mathit{SB}, 1, !\top, \{t_B,\mathit{info_B}\})\)};
\node[sys,win, below=of 2, dashed] 		(4)	{\(\mathit{EA}\),\\\((\mathit{SAA}, 1, \top, \{\})\),\\\((\mathit{SB}, 1, !\top, \{t_B,\mathit{info_B}\})\)};
\node[sys,win, below=of 4, dashed] 		(6)	{\(\mathit{EA}\),\\\((\mathit{SAA}, 1, !\top, \{\mathit{info_B}\})\),\\\((\mathit{SB}, 1, !\top, \{t_B,\mathit{info_B}\})\)};
\node[sys,win, below=of 6,dashed] 		(8)	{\(\mathit{EA}\),\\\((\mathit{pA}, 1, !\top, \{\mathit{aa}\})\),\\\((\mathit{pB}, 1, !\top, \{\mathit{ba}\})\)};
\node[sys,win, below=of 8] 		(10)	{\(\mathit{EA}\),\\\((\mathit{AA}, 1, !\top, \{\})\),\\\((\mathit{pB}, 1, !\top, \{\mathit{ba}\})\)};
\node[env,win, below=of 10] 		(12)	{\(\mathit{EA}\),\\\((\mathit{AA}, 1, !\top, \{\})\),\\\((\mathit{BA}, 1, !\top, \{\})\)};

\node[env,win, below=of 0,right=of 2] 		(1)	{\(\mathit{LB}\),\\\((\mathit{SA}, 1, !\top, \{t_A,\mathit{info_A}\})\),\\\((\mathit{SB}, 1, !\top, \{t_B,\mathit{info_B}\})\)};
\node[sys,win, below=of 1, dashed] 		(3)	{\(\mathit{EB}\),\\\((\mathit{SA}, 1, !\top, \{t_A,\mathit{info_A}\})\),\\\((\mathit{SBB}, 1, \top, \{\})\)};
\node[sys,win, below=of 3, dashed] 		(5)	{\(\mathit{EB}\),\\\((\mathit{SA}, 1, !\top, \{t_A,\mathit{info_A}\})\),\\\((\mathit{SBB}, 1, !\top, \{\mathit{info_A}\})\)};
\node[sys,win, below=of 5, dashed] 		(7)	{\(\mathit{EB}\),\\\((\mathit{pA}, 1, !\top, \{\mathit{ab}\})\),\\\((\mathit{pB}, 1, !\top, \{\mathit{bb}\})\)};
\node[sys,win, below=of 7] 		(9)	{\(\mathit{EB}\),\\\((\mathit{AB}, 1, !\top, \{\})\),\\\((\mathit{pB}, 1, !\top, \{\mathit{bb}\})\)};
\node[env,win, below=of 9] 		(11)	{\(\mathit{EB}\),\\\((\mathit{AB}, 1, !\top, \{\})\),\\\((\mathit{BB}, 1, !\top, \{\})\)};

\node[sys, left=of 0, dashed]		(fa0)	{\(\mathit{Env}\),\\\((\mathit{SA}, 1, !\top, \{\mathit{fa}_A\})\),\\\((\mathit{SB}, 1, !\top, \{t_B,\mathit{info_B}\})\)};
\node[sys, badstate, bottom color=black, top color=black,above=of fa0] 	(fa1E)	{\(\mathit{Env}\),\\\((\mathit{pA}, 1, !\top, \{\mathit{aa},\mathit{ab}\})\),\\\((\mathit{SB}, 1, !\top, \{t_B,\mathit{info_B}\})\)};

\node[sys, dashed, below=of fa0] 	(fa1)	{\(\mathit{Env}\),\\\((\mathit{pA}, 1, !\top, \{\mathit{aa}\})\),\\\((\mathit{SB}, 1, !\top, \{t_B,\mathit{info_B}\})\)};
\node[sys, dashed,label={[yshift=-5mm, xshift=-5mm]above left:\circled{2}}, below=of fa1] 	(fa2a)	{\(\mathit{Env}\),\\\((\mathit{AA}, 1, !\top, \{t_\bot\})\),\\\((\mathit{SB}, 1, !\top, \{t_B,\mathit{info_B}\})\)};
\node[env, below=of fa2a] 	(fa2b)	{\(\mathit{Env}\),\\\((\mathit{AA}, 1, !\top, \{\})\),\\\((\mathit{SB}, 1, !\top, \{t_B,\mathit{info_B}\})\)};
\node[env, below=of fa2b] 	(fa4)	{\(\mathit{LA}\),\\\((\mathit{AA}, 1, !\top, \{\})\),\\\((\mathit{SB}, 1, !\top, \{t_B,\mathit{info_B}\})\)};
\node[env, below=of fa4] 	(fa3)	{\(\mathit{LB}\),\\\((\mathit{AA}, 1, !\top, \{\})\),\\\((\mathit{SB}, 1, !\top, \{t_B,\mathit{info_B}\})\)};
\node[sys, dashed,label={[yshift=-5mm, xshift=0mm]above left:\circled{3}},below=of fa3] 	(fa5)	{\(\mathit{EB}\),\\\((\mathit{AA}, 1, !\top, \{\})\),\\\((\mathit{SBB}, 1, \top, \{\})\)};
\node[env, right=of init, xshift=13mm] 	(bad1)	{\(\mathit{Env}\),\\\((\mathit{SA}, 1, !\top, \{\})\),\\\((\mathit{SB}, 1, !\top, \{\})\)};
\node[env, badstate, below=of bad1](bad2)	{\(\mathit{LA}\),\\\((\mathit{SA}, 1, !\top, \{\})\),\\\((\mathit{SB}, 1, !\top, \{\})\)};
\node[env, badstate, right=of bad1, xshift=10mm](bad3)	{\(\mathit{LB}\),\\\((\mathit{SA}, 1, !\top, \{\})\),\\\((\mathit{SB}, 1, !\top, \{\})\)};
\node[env, below=of bad3, dashed] 	(fr1)	{\(\mathit{Env}\),\\\((\mathit{SA}, 1, !\top, \{\mathit{info}_A\})\),\\\((\mathit{SB}, 1, !\top, \{\mathit{t_B}\})\)};
\node[env, below=of fr1]	 	(fr2)	{\(\mathit{LB}\),\\\((\mathit{SA}, 1, !\top, \{\mathit{info}_A\})\),\\\((\mathit{SB}, 1, !\top, \{\mathit{t_B}\})\)};
\node[sys, below=of fr2, dashed] 	(fr3cor){\(\mathit{EB}\),\\\((\mathit{SA}, 1, !\top, \{\mathit{info}_A\})\),\\\((\mathit{SBB}, 1, \top, \{\})\)};
\node[sys, below=of fr3cor, dashed,label={[yshift=-5mm, xshift=-3.5mm]above left:\circled{4}}] 	(fr3)	{\(\mathit{EB}\),\\\((\mathit{SA}, 1, !\top, \{\mathit{info}_A\})\),\\\((\mathit{SBB}, 1, !\top, \{\mathit{fr_B}\})\)};

\draw[->,win, >=stealth']
   (init) edge (0)
   (9) edge node[right, xshift=2mm] {\(\mathit{bb}\)}	 (11)
   (5) edge node[right, xshift=2mm] {\(\mathit{info_A}\)}	 (7)
   (0) edge node[below left, pos=0.1, xshift=4mm] {\(i_B\)} 	(1)
   (1) edge node[right, xshift=2mm] {\(t_B\)} 	(3)
   (0) edge node[left, xshift=-2mm] {\(i_A\)} 	(2)
   (2) edge node[left, xshift=-2mm] {\(t_A\)} 	(4)
   (4) edge (6)
   (6) edge node[left, xshift=-2mm] {\(\mathit{info_B}\)} 	(8)
   (7) edge node[right, xshift=2mm] {\(\mathit{ab}\)} 	(9)
   (3) edge (5)
   (8) edge node[left, xshift=-2mm] {\(\mathit{aa}\)} 	(10)
   (10) edgenode[left, xshift=-2mm] {\(\mathit{ba}\)} 	 (12)
;
\draw[->, >=stealth']
   (init) edge (fa0)
   (fa0) edge node[left, xshift=-2mm] {\(\mathit{fa}_A\)}	 (fa1)
   (fa0) edge node[left, xshift=-2mm] {\(\mathit{fa}_A\)}	 (fa1E)
   (fa1) edge node[left, xshift=-2mm] {\(\mathit{aa}\)}	 	(fa2a)
;
\draw[->, >=stealth']
 (fa1.west) -- ([xshift=-2mm]fa1.west) -- node[left] {\(\mathit{aa}\)}  ([xshift=-2mm, yshift=3mm]fa2b.west) -- 	([yshift=3mm]fa2b.west)
;

\draw[->, >=stealth']
 ([yshift=-3mm]fa2b.west) -- ([xshift=-2mm,yshift=-3mm]fa2b.west) -- node[left] {\(\mathit{i_B}\)}  ([xshift=-2mm]fa3.west) -- 	(fa3.west)
;
\draw[->, >=stealth']
   (fa2b) edge node[left, xshift=-2mm] {\(\mathit{i_A}\)} 	(fa4)
   (fa3) edge node[left, xshift=-2mm] {\(\mathit{t_B}\)} 	(fa5)
   (init) edge (bad1)
   (bad1) edge node[left, xshift=0mm] {\(\mathit{i_A}\)} 	(bad2)
   (bad1) edge node[above, xshift=0mm] {\(\mathit{t_B}\)} 	(bad3)
;
\draw[->, >=stealth'] (init.north) -- ([yshift=2mm]init.north) -- ([xshift=2mm,yshift=10mm]bad3.east) -- ([xshift=2mm,yshift=-10mm]bad3.east) -- ([yshift=5mm]fr1.north) -- (fr1.north);

\draw[->, >=stealth']
   (fr1) edge node[right, xshift=0mm] {\(\mathit{i_B}\)} 	(fr2)
   (fr2) edge node[right, xshift=0mm] {\(\mathit{t_B}\)} 	(fr3cor)
   (fr3cor) edge (fr3)
;

\node [below=of 12, xshift=20mm, yshift=5mm] (al) {\color{orange}Winning strategy for the system players};

\begin{pgfonlayer}{background}
\draw [-, rectangle,rounded corners,orange,fill=orange!15] ([xshift=-8mm,yshift=3mm]init.north west) -- ([xshift=9mm,yshift=3mm]init.north east) --([xshift=2.4mm,yshift=-5mm]0.east) -- ([xshift=-10mm,yshift=2mm]1.north) -- ([xshift=1mm,yshift=2mm]1.north east) -- ([xshift=7mm,yshift=-10mm]11.south east) -- ([xshift=-7mm,yshift=-10mm]12.south west) -- cycle;
\end{pgfonlayer}
\end{tikzpicture}

%% file: figures/securitySystemBoundedUnfolding.tex
\begin{tikzpicture}[node distance=1.25cm,>=stealth',bend angle=45,auto,scale=0.8]
	\node [envplace,tokens=1] (env)  [label=below:$\mathit{Env}$, tokens=1]                                  {};
	\node [transition] (eL)  [left of=env, xshift=-5mm,label={[label distance=-0.5mm]below:$i_A$}] {};
	\node [transition] (eR)  [right of=env, xshift=5mm,label={[label distance=-0.5mm]below:$i_B$}] {};
	\node [envplace] (epL)  [label=left:$\mathit{LA}$, left of=eL, xshift=-5mm]                                  {};
	\node [envplace] (epR)  [label=right:$\mathit{LB}$, right of=eR, xshift=5mm]                                  {};
	\node [transition] (tA)  [below of=epL,label={[label distance=-1mm]above right:$t_A$}] {};
	\node [sysplace] (IA) [below of=tA,label=left:$\mathit{SAA}$]                     {};
	\node [envplace] (ea)  [left of=tA,xshift=3.6mm,label=above:$\mathit{EA}$]                                  {};
	\node [sysplace] (A) [right of=tA,label=right:$\mathit{SA}$,tokens=1]                     {};
	\node [transition] (t2)  [below of=IA,xshift=2mm,label=right:\(\mathit{fa}_A\)] {};
	\node [transition] (t1)  [left of=t2,xshift=1.6mm,label={[label distance=-1mm]left:\(\mathit{fr}_A\)}] {};
	\node [transition] (t3)  [right of=t2,xshift=13mm,yshift=5mm,label=above:$\mathit{info}_B$] {};
	\node [sysplace] (pa)  [below of=t2,xshift=-6mm,label={[label distance=-1mm]left:$\mathit{pA}'$}]                     {};
	\node [sysplace] (pa1) [below of=t1,xshift=-10mm,label={[label distance=-1mm]left:$\mathit{pA}$}]                     {};
	\node [sysplace] (pa2) [right of=pa,xshift=3mm,label={[label distance=-1mm]left:$\mathit{pA}_A$}]                     {};
	\node [sysplace] (pa3) [right of=pa2,xshift=4mm,label={[label distance=-1mm]left:$\mathit{pA}_B$}]                     {}; 
	\node [transition] (tAB)  [below of=pa,xshift=2.5mm] {};
	\node [transition] (tAA)  [below of=pa,xshift=-2.5mm] {};
	\node [transition] (tAB1)  [below of=pa1,xshift=2.5mm] {};
	\node [transition] (tAA1)  [below of=pa1,xshift=-2.5mm] {};
	\node [transition] (tAB2)  [below of=pa2,xshift=2.5mm] {};
	\node [transition] (tAA2)  [below of=pa2,xshift=-2.5mm] {};
	\node [transition] (tAB3)  [below of=pa3,xshift=2.5mm] {};
	\node [transition] (tAA3)  [below of=pa3,xshift=-2.5mm] {};
	\node [sysplace] (ab) [below of=pa2, yshift=-1.25cm,label=below:$\mathit{AB}$]                     {};
	\node [sysplace] (aa) [below of=pa, yshift=-1.25cm,label=below:$\mathit{AA}$]                     {};
	\node [transition] (tB)  [below of=epR,label={[label distance=-1mm]above left:$t_B$}] {};
	\node [sysplace] (IB) [below of=tB,label=right:$\mathit{SBB}$]                     {};
	\node [envplace] (eb)  [right of=tB,xshift=-3.6mm,label=above:$\mathit{EB}$]                                  {};
	\node [sysplace] (B) [left of=tB,label=left:$\mathit{SB}$,tokens=1]                     {};
	\node [transition] (t2b)  [below of=IB,xshift=-2mm,label=left:\(\mathit{fa}_B\)] {};
	\node [transition] (t1b)  [right of=t2b,xshift=-1.6mm,label={[label distance=-1mm]right:\(\mathit{fr}_B\)}] {};
	\node [transition] (t3b)  [left of=t2b,xshift=-13mm, yshift=5mm,label=above:$\mathit{info}_A$] {};
	\node [sysplace] (pb)  [below of=t2b,xshift=0mm,xshift=6mm,label={[label distance=-1mm]right:$\mathit{pB}'$}]                     {};
	\node [sysplace] (pb1) [below of=t1b,xshift=0mm,xshift=10mm,label={[label distance=-1mm]right:$\mathit{pB}$}]                     {};
	\node [sysplace] (pb2) [left of=pb,xshift=-3mm,label={[label distance=-1mm]right:$\mathit{pB}_B$}]                     {};
	\node [sysplace] (pb3) [left of=pb2,xshift=-4mm,label={[label distance=-1mm]right:$\mathit{pB}_A$}]                     {}; 
	\node [transition] (tBB)  [below of=pb,xshift=2.5mm] {};
	\node [transition] (tBA)  [below of=pb,xshift=-2.5mm] {};
	\node [transition] (tBB1)  [below of=pb1,xshift=2.5mm] {};
	\node [transition] (tBA1)  [below of=pb1,xshift=-2.5mm] {};
	\node [transition] (tBB2)  [below of=pb2,xshift=2.5mm] {};
	\node [transition] (tBA2)  [below of=pb2,xshift=-2.5mm] {};
	\node [transition] (tBB3)  [below of=pb3,xshift=2.5mm] {};
	\node [transition] (tBA3)  [below of=pb3,xshift=-2.5mm] {};
	\node [sysplace] (bb) [below of=pb, yshift=-1.25cm,label=below:$\mathit{BB}$]                     {};
	\node [sysplace] (ba) [below of=pb2, yshift=-1.25cm,label=below:$\mathit{BA}$]                     {};

\node [right=of A, yshift=-4mm, xshift=-2mm] (intr) {\DarkBlue{}Burglary};
\node [below=of t3, xshift=7mm, yshift=10mm] (com) {\color{ganttGreen}Comm.};
\node [right=of aa, xshift=16.5mm, yshift=-7mm] (al) {\Red{}Alarm};

	\path[-latex, thick] 	(eL)  		edge [pre]                            (env)
					edge [post]                            (epL)
		 	(tA)  		edge [pre]                            (epL)
		 	  		edge [pre]                            (A)
					edge [post]                            (ea)
					edge [post]                            (IA)
			(t1)  		edge [pre]                            (IA)
					edge [post]                            (pa1)
			(t2)  		edge [pre]                            (A)
					edge [post]                            (pa)
			(t3)  		edge [pre]                            (IA)
			  		edge [pre]                            (B)
					edge [post]                            (pa2)
					edge [post]                            (pb3)
			(tAB)  		edge [pre]                            (pa)
					edge [post]                            (ab)
			(tAA)  		edge [pre]                            (pa)
					edge [post]                            (aa)
			(tAB1)  		edge [pre]                            (pa1)
					edge [post]                            (ab)
			(tAA1)  		edge [pre]                            (pa1)
					edge [post]                            (aa)
			(tAB2)  		edge [pre]                            (pa2)
					edge [post]                            (ab)
			(tAA2)  		edge [pre]                            (pa2)
					edge [post]                            (aa)
			(tAB3)  		edge [pre]                            (pa3)
					edge [post]                            (ab)
			(tAA3)  		edge [pre]                            (pa3)
					edge [post]                            (aa)				
	;
	\path[->, thick] 	(eR)  		edge [pre]                            (env)
					edge [post]                            (epR)
		 	(tB)  		edge [pre]                            (epR)
		 	  		edge [pre]                            (B)
					edge [post]                            (eb)
					edge [post]                            (IB)
			(t1b)  		edge [pre]                            (IB)
					edge [post]                            (pb1)
			(t2b)  		edge [pre]                            (B)
					edge [post]                            (pb)
			(t3b)  		edge [pre]                            (IB)
			  		edge [pre]                            (A)
					edge [post]                            (pb2)
					edge [post]                            (pa3)
			(tBB)  		edge [pre]                            (pb)
					edge [post]                            (bb)
			(tBA)  		edge [pre]                            (pb)
					edge [post]                            (ba)
			(tBB1)  		edge [pre]                            (pb1)
					edge [post]                            (bb)
			(tBA1)  		edge [pre]                            (pb1)
					edge [post]                            (ba)
			(tBB2)  		edge [pre]                            (pb2)
					edge [post]                            (bb)
			(tBA2)  		edge [pre]                            (pb2)
					edge [post]                            (ba)
			(tBB3)  		edge [pre]                            (pb3)
					edge [post]                            (bb)
			(tBA3)  		edge [pre]                            (pb3)
					edge [post]                            (ba)		
		;

\begin{pgfonlayer}{background}
\draw [-, rectangle,rounded corners,DarkBlue,fill=cdcBlueL!15] ([xshift=-2mm,yshift=18mm]ea.north west) -- ([xshift=2mm,yshift=18mm]eb.north east) -- ([xshift=2mm,yshift=-5mm]eb.south east) -- ([xshift=-2mm,yshift=-5mm]ea.south west) -- cycle;
\draw [-, rectangle,rounded corners,DarkRed,fill=DarkRed!15, fill opacity=0.95] ([xshift=-10mm,yshift=4mm]pa1.north west) -- ([xshift=10mm,yshift=4mm]pb1.north east) -- ([xshift=10mm,yshift=-40mm]pb1.south east) -- ([xshift=-10mm,yshift=-40mm]pa1.south west) -- cycle;
\draw [-, rectangle,rounded corners,ganttGreen,fill=cdcGreenL!15, fill opacity=0.45] ([xshift=-5mm,yshift=-10mm]t3.south west) -- ([xshift=-5mm,yshift=10mm]t3.north west) -- ([xshift=5mm,yshift=10mm]t3b.north east) -- ([xshift=5mm,yshift=-10mm]t3b.south east) -- cycle;
\end{pgfonlayer}
\end{tikzpicture}

%% file: figures/benchmarksMerged.tex
{\scriptsize 
\begin{longtable}{l|c|ccc|Hcc|cc|cc|cc|cc} 
\hline\hline\noalign{\smallskip}
 & & & & & \multicolumn{4}{c}{\hspace{0.95cm}\emph{Symbolic Synthesis}} && \multicolumn{5}{c}{\hspace{0.4cm}\emph{Bounded Synthesis}}\\
\it{Ben.} & \it{Par.} & $\#\mathit{Tok}$ & $\#\pl$ & $\#\tr$ & $\#\mathit{Var}$ & \emph{time} & \emph{memory} & $\#\pl_\mathit{str}
$ & $\#\tr_{\mathit{str}}$ & $n$ & $b$ & \emph{time} & \emph{memory} & $\#\pl_\mathit{str}$ & $\#\tr_\mathit{str}$\\
\noalign{\smallskip}
\hline
\noalign{\smallskip}
AS	&   2  & 3 & 17 & 26  & 78   &\bf 1.9 &  .30   &\bf 17&\bf 10&7 & 2 & 18.0 &\bf .18 &\bf 17 &\bf 10 \\
	&   3  & 4 & 28 & 69  & 180  &  2.5   &  .41   &  31 & 18   & 7 & 3 & \multicolumn{4}{l}{timeout}\\
    &   &\multicolumn{2}{c}{\hspace{1cm}\dots} && \multicolumn{4}{c}{\hspace{1.5cm}\dots} && \multicolumn{3}{c}{\hspace{-0.5cm}\dots}\\
	&   6  & 7 & 73 & 462 & 1052 &  91.0  &  4.65  &  97 & 54   & 7 & 6 & \multicolumn{4}{l}{timeout}\\
	&   7  &   &    &     &      &  \multicolumn{3}{l}{timeout}&& 7 & 7 & \multicolumn{4}{l}{timeout}\\
\hline

CM &   2/1  & 6  & 13 & 10  & 66  &  1.4 &  .30   &  14 &\bf 8  & 6 & 3 &\bf .6  &\bf .06  &\bf 13 &\bf 8 \\
   &   2/2  & 7  & 18 & 16  & 96  &  2.0  &  .30   &  -  & -     & - & - & \\
    &   &\multicolumn{2}{c}{\hspace{1cm}\dots} && \multicolumn{4}{c}{\hspace{1.5cm}\dots} && \multicolumn{3}{c}{\hspace{-0.5cm}\dots}\\
   &   2/5  & 10 & 33 & 34  & 186 &  50.9 &  2.74  &  -  & -     & - & - & \\
   &   2/6  &    &    &     &     &  \multicolumn{3}{l}{timeout}&& - & - & \\
\cline{2-16}
   &   3/1  & 8  & 18 & 15  & 92  &  2.0  &  .30   &  26 & 12    & 6 & 3 &\bf 1.7 &\bf .12  &\bf 18 &\bf 9 \\
   &   3/2  & 9  & 25 & 24  & 132 &  2.4  &  .30   &  36 & 18    & 6 & 4 & \multicolumn{4}{l}{timeout}\\
   &   3/3  & 10 & 32 & 33  & 172 &  3.8  &  .39   &  -  & -     & - & - & \\
   &   3/4  & 11 & 39 & 42  & 212 &  17.2 &  1.28  &  -  & -     & - & - & \\
   &   3/5  &    &    &     &     &  \multicolumn{3}{l}{timeout}&& \\ 
\cline{2-16}
   &   4/1  & 10 & 23 & 20  & 120 &\bf 2.3&  .30  &  42 & 16         & 6 & 3 &  6.0 &\bf .19  &\bf 21&\bf 12\\
   &   4/2  & 11 & 32 & 32  & 172 &  5.0  &  .40  &  55 & 24         & 6 & 4 & \multicolumn{4}{l}{timeout}\\
   &   4/3  & 12 & 41 & 44  & 224 &  10.9 &  .84  &  68 & 32         & 6 & 5 & \multicolumn{4}{l}{timeout}\\
   &   4/4  & 13 & 50 & 56  & 276 &  92.2 &  4.17 &  -  & -          & - & - & \\ 
   &   4/5  &    &    &     &     &  \multicolumn{3}{l}{out of memory}&& - & - & \\ 
\cline{2-16}
   &   5/1  & 12 & 28 & 25  & 146 &\bf 7.2    &  .39   &  62  & 20     & 6 & 3 & 11.1 &\bf .19  &\bf 22 &\bf 11 \\
   &   5/2  & 13 & 39 & 40  & 208 &  20.8   &  .79   &  78  & 30     & 6 & 4 & \multicolumn{4}{l}{timeout}\\
   &   5/3  & 14 & 50 & 55  & 270 &  82.1   &  2.67  &  94  & 40     & 6 & 5 & \multicolumn{4}{l}{timeout}\\
   &   5/4  & 15 & 61 & 70  & 332 &  1101.3 &  16.70 &  110 & 50     & 6 & 6 & \multicolumn{4}{l}{timeout}\\
   &   5/5  &    &    &     &     &  \multicolumn{3}{l}{out of memory}&& - & - & \\ 
\cline{2-16}
   &   6/1  & 14 &  33 & 30  & 172 & 41.5   &  .80   &  86  & 24     & 6 & 3 &\bf 23.6 &\bf .31  &\bf 25 &\bf 12 \\
   &   6/2  & 15 &  46 & 48  & 244 & 183.4  &  2.67  &  105 & 36     & 6 & 4 & \multicolumn{4}{l}{timeout}\\
\cline{2-16}
   &   7/1  & 16 &  38 & 35  & 198 & 289.5  &  5.35  &  114 & 28     & 6 & 3 &\bf 26.0 &\bf .36  &\bf 27 &\bf 13 \\
\cline{2-16}
   &   8/1  & 18 & 43  & 40  & 226 & 1657.4  &  15.73  &  146 & 32   & 6 & 3 &\bf 94.7 &\bf .65  &\bf 27 &\bf 14 \\
\cline{2-16}
   &   9/1  & 20 & 48  & 45  &     &  \multicolumn{3}{l}{out of memory} && 6 & 3 &  152.4  &  1.22  &  36 & 25 \\
\cline{2-16}
    &   &\multicolumn{2}{c}{\hspace{1cm}\dots} && \multicolumn{4}{c}{\hspace{1.5cm}\dots} && \multicolumn{3}{c}{\hspace{-0.5cm}\dots}\\
\cline{2-16}
   &   15/1   & 32& 78  & 75  &     &  \multicolumn{3}{l}{out of memory} && 6 & 3 & 1259.5  &  23.24  &  66 & 49 \\
\hline

SR &   2/1  & 5 & 18 & 17  & 86 &  \bf1.9  &  .30   &  32 & 16    & 6 & 2 & 2.7  &\bf .16  & \bf 18 &\bf 10 \\
   &   2/2  & 6 & 24 & 26  &116 &     4.3  &  .39   &   - & -     & - & - & \\
   &   2/3  & 7 & 30 & 35  &144 &  1290.3  &  5.36  &   - & -     & - & - & \\
   &   2/4  &   &    &     &    &  \multicolumn{3}{l}{out of memory}&& - & - & \\
\cline{2-16}
   &   3/1  &   &    &     &    &  \multicolumn{3}{l}{out of memory}&&7 & 2 &  219.4  &  .61  &  27& 19 \\ 
\hline

JP &   2  & 3  & 12  & 13  & 46   &\bf 1.3 &  .30   &\bf 16&\bf 13 & 7 & 3 &  1.5  & \bf .08  &\bf 16&\bf 13\\
   &   3  & 4  & 18  & 23  & 76   &  1.8   &  .30   &  34  & 28  & 8 & 4 & \multicolumn{4}{l}{timeout}\\
    &   &\multicolumn{2}{c}{\hspace{1cm}\dots} && \multicolumn{4}{c}{\hspace{1.5cm}\dots} && \multicolumn{3}{c}{\hspace{-0.5cm}\dots}\\
   &   11 & 12 & 102 & 175 & 762  &  353.3 &  16.78 &  706 & 484 &16 &12 & \multicolumn{4}{l}{timeout}\\
   &   12 &    &     &     &      &  \multicolumn{3}{l}{out of memory}&&17 &13 & \multicolumn{4}{l}{timeout}\\ 
\hline

DW &   1   & 3  & 12  & 10  & 46  &  1.1    &  .30   &\bf10  &\bf 6&  8 & 1 & \bf.3  &\bf.04   &\bf10 &\bf6 \\
   &   2   & 4  & 19  & 16  & 72  &  2.0    &  .30   &  24   & 16  & 10 & 1 & \bf.4  &\bf.05   &\bf16 &\bf12\\
    &   &\multicolumn{2}{c}{\hspace{1cm}\dots} && \multicolumn{4}{c}{\hspace{1.5cm}\dots} && \multicolumn{3}{c}{\hspace{-0.5cm}\dots}\\
   &   11  & 13 & 82  & 70  & 296 &  137.6  &  2.82  &  420  & 286 & 28 & 1 &\bf78.3 &\bf2.26  &\bf70 &\bf57\\
   &   12  & 14 & 89  & 76  & 320 &  201.8  &\bf2.82 &  494  & 336 & 30 & 1 &\bf157.3&   3.38  &\bf76 &\bf62\\
   &   13  & 15 & 96  & 82  & 344 &\bf277.9 &\bf4.24 &  574  & 390 & 32 & 1 & 341.1  &  8.18   &\bf82 &\bf67\\
   &   14  & 16 & 103 & 88  & 368 &\bf400.1 &\bf4.22 &  660  & 448 & 34 & 1 & 624.5  &  11.80  &\bf88 &\bf72\\
   &   15  & 17 & 110 & 94  & 394 &  537.2  &  4.87  &  752  & 510 & 36 & 1 & \multicolumn{4}{l}{timeout}\\
    &   &\multicolumn{2}{c}{\hspace{1cm}\dots} && \multicolumn{4}{c}{\hspace{1.5cm}\dots} && \multicolumn{3}{c}{\hspace{-0.5cm}\dots}\\
   &   20  & 22 & 145 & 124 & 516 &  1799.8 &  11.69 &  1302 & 880 & 46 & 1 & \multicolumn{4}{l}{timeout}\\
   &   21  &    &     &     &     &  \multicolumn{3}{l}{timeout}&  & 48 & 1 & \multicolumn{4}{l}{timeout}\\ 
\hline

DWs &   1   & 3  & 11  & 6   & 36  &  .7     &  .29   &\bf8   &\bf3 &  5 & 1 &\bf.2    &\bf.04   &\bf8  &\bf3 \\
    &   2   & 5  & 21  & 12  & 70  &  1.6    &  .30   &  23   & 10  &  7 & 1 &\bf.3    &\bf.05   &\bf17 &\bf8 \\
    &   &\multicolumn{2}{c}{\hspace{1cm}\dots} && \multicolumn{4}{c}{\hspace{1.5cm}\dots} && \multicolumn{3}{c}{\hspace{-0.5cm}\dots}\\
    &   7   & 15 & 71  & 42  & 232 &  14.4   &  .91   &  218  & 105 & 17 & 1 &\bf5.3   &\bf.87   &\bf57 &\bf28\\
    &   8   & 17 & 81  & 48  & 266 &  24.9   &\bf1.52 &  281  & 136 & 19 & 1 &\bf11.7  &   3.18  &\bf65 &\bf32\\
    &   9   & 19 & 91  & 54  & 298 &  45.4   &\bf2.83 &  352  & 171 & 21 & 1 &\bf41.3  &   8.84  &\bf73 &\bf36\\
    &   10  & 21 & 101 & 60  & 330 &\bf80.0  &\bf2.85 &  431  & 210 & 23 & 1 &  132.8  &   12.96 &\bf81 &\bf40\\
    &   11  & 23 & 111 & 66  & 362 &  142.6  &  4.42  &  518  & 253 & 25 & 1 & \multicolumn{4}{l}{timeout}\\
    &   &\multicolumn{2}{c}{\hspace{1cm}\dots} && \multicolumn{4}{c}{\hspace{1.5cm}\dots} && \multicolumn{3}{c}{\hspace{-0.5cm}\dots}\\
    &   16  & 33 & 161 & 96  & 524 &  1508.3 &  16.30 &  1073 & 528 & 35 & 1 & \multicolumn{4}{l}{timeout}\\
    &   17  &    &      &    &      &  \multicolumn{3}{l}{timeout}&&  37 & 1 & \multicolumn{4}{l}{timeout}\\ 
\hline

& \multicolumn{15}{l}{'-' means no winning strategy exists.}\\
\end{longtable}
}

%% file: figures/tableVariableSize.tex
{

\begin{longtable}{l|c|c|cc|cccc}
\hline\hline\noalign{\smallskip}
\it{Ben.} & \it{Par.} & $\#\mathit{Var}_\mathit{symbolic}$ & n & b & $\#\mathit{Var}_\mathit{bounded}$ & $\#\mathit{Var}_\exists$ & $\#\mathit{Var}_\forall$ & $\#\mathit{Var}_{\phi_n}$\\
\noalign{\smallskip}
\hline
\noalign{\smallskip}
CM &   2/1  & 66  &  6 & 3 &  2743 &  53  & 162 & 2528\\
   &   3/1  & 92  &  6 & 3 &  7678 & 109  & 300 & 7269\\
   &   4/1  & 120 &  6 & 3 & 17849 & 197  & 462 & 17190\\
   &   5/1  & 146 &  6 & 3 & 25848 & 266  & 570 & 25012\\
   &   6/1  & 172 &  6 & 3 & 35287 & 343  & 678 & 34266\\
   &   7/1  & 198 &  6 & 3 & 46166 & 428  & 786 & 44952\\
   &   8/1  & 226 &  6 & 3 & 58485 & 521  & 894 & 57070\\
\hline

DW &   1   & 46  &    8 & 1 &  1144 &  12 &   96 & 1036\\
   &   2   & 72  &   10 & 1 &  2591 &  20 &  190 & 2381\\
   &   3   & 98  &   12 & 1 &  4838 &  28 &  312 & 4498\\
   &   4   & 124 &   14 & 1 &  8052 &  36 &  462 & 7554\\
   &   5   & 148 &   16 & 1 & 12404 &  44 &  640 & 11720\\
   &   6   & 172 &   18 & 1 & 18059 &52   &  846 & 17161\\
   &   7   & 198 &   20 & 1 & 25186 &60   & 1080 & 24046\\
   &   8   & 224 &   22 & 1 & 33953 &68   & 1342 & 32543\\
   &   9   & 248 &   24 & 1 & 44528 &76   & 1632 & 42820\\
   &   10  & 272 &   26 & 1 & 57079 &84   & 1950 & 55045\\
   &   11  & 296 &   28 & 1 & 71744 &92   & 2296 & 69356\\
   &   12  & 320 &   30 & 1 & 88781 &100  & 2670 & 86011\\
   &   13  & 344 &   32 & 1 &108268 &108  & 3072 & 105088\\
   &   14  & 368 &   34 & 1 &130403 &116  & 3502 & 126785\\
\hline

DWs &   1   & 36  &  5 & 1 &   440&9    &  55 & 376\\
    &   2   & 70  &  7 & 1 &  1414&17   & 147 & 1250\\
    &   3   & 102 &  9 & 1 &  3204&25   & 279 & 2900\\
    &   4   & 136 & 11 & 1 &  6050&33   & 451 & 5566\\
    &   5   & 168 & 13 & 1 & 10192&41   & 663 & 9488\\
    &   6   & 200 & 15 & 1 & 15870&49   & 915 & 14906\\
    &   7   & 232 & 17 & 1 & 23324&57   &1207 & 22060\\
    &   8   & 266 & 19 & 1 & 32794&   65&1539 & 31190\\
    &   9   & 298 & 21 & 1 & 44520&   73&1911 & 42536\\
    &   10  & 330 & 23 & 1 & 58742&   81&2323 & 56338\\
\hline
\end{longtable}
}